\documentclass[fleqn,10pt]{wlscirep}

\usepackage[utf8]{inputenc}
\usepackage[T1]{fontenc}
\usepackage{geometry}
\usepackage{indentfirst}
\usepackage{enumitem}

\usepackage{amsmath, amssymb, amsfonts, amsthm, mathrsfs}
\usepackage{siunitx}

\usepackage{algorithm}
\usepackage{algorithmicx}
\usepackage{algpseudocode}

\usepackage{tabularx}
\usepackage{booktabs}
\usepackage{multirow}
\usepackage{makecell}
\usepackage{arydshln}
\usepackage{nicematrix}

\usepackage{graphicx}
\usepackage{float}
\usepackage{wrapfig}
\usepackage{subcaption}
\usepackage{subcaption}
\usepackage[font=small,labelfont=bf]{caption}
\usepackage{overpic}

\usepackage{microtype}
\usepackage{nicefrac}
\usepackage{anyfontsize}
\usepackage{textcomp}
\usepackage{ulem}

\usepackage{xargs}

\usepackage{lineno}
\usepackage{listings}
\usepackage{lipsum}
\usepackage{mdframed}
\usepackage[title]{appendix}
\usepackage{url}
\usepackage[numbers]{natbib}
\usepackage{hyperref}
\usepackage{manyfoot}

\newcounter{hypothesiscounter}
\newcommand*{\dslabel}[1]{DS\refstepcounter{hypothesiscounter}\thehypothesiscounter\label{#1}}
\newcommand*{\dsref}[1]{DS\ref{#1}}

\definecolor{codegreen}{rgb}{0,0.6,0}
\definecolor{codegray}{rgb}{0.5,0.5,0.5}
\definecolor{codepurple}{rgb}{0.58,0,0.82}
\definecolor{backcolour}{rgb}{0.95,0.95,0.92}

\title{SuryaBench: Benchmark Dataset for Advancing Machine Learning in Heliophysics and Space Weather Prediction}

\author[1,2,*]{Sujit Roy}
\author[3,4]{Dinesha V. Hegde}
\author[5]{Johannes Schmude}
\author[1]{Amy Lin}
\author[1]{Vishal Gaur}
\author[1]{Rohit Lal}
\author[1]{Kshitiz Mandal}
\author[6]{Talwinder Singh}
\author[7]{Andr\'es Mu\~noz-Jaramillo}
\author[6]{Kang Yang}
\author[6]{Chetraj Pandey}
\author[6]{Jinsu Hong}
\author[6]{Berkay Aydin}
\author[8]{Ryan McGranaghan}
\author[9]{Spiridon Kasapis}
\author[10]{Vishal Upendran}
\author[11]{Shah Bahauddin}
\author[12]{Daniel da Silva}
\author[5]{Marcus Freitag}
\author[1]{Iksha Gurung}
\author[3,4]{Nikolai Pogorelov}
\author[5]{Campbell Watson}
\author[2]{Manil Maskey}
\author[5]{Juan Bernabe-Moreno}
\author[2]{Rahul Ramachandran}

\affil[1]{Earth System Science Center, University of Alabama in Huntsville, AL, USA}
\affil[2]{NASA Marshall Space Flight Center, Huntsville, AL, USA}
\affil[3]{Department of Space Science, The University of Alabama in Huntsville, AL, USA}
\affil[4]{Center for Space Plasma and Aeronomic Research (CSPAR), The University of Alabama in Huntsville, AL, USA}
\affil[5]{IBM Research}
\affil[6]{Georgia State University}
\affil[7]{Southwest Research Institute}
\affil[8]{NASA Jet Propulsion Laboratory}
\affil[9]{Princeton University}
\affil[10]{SETI Institute}
\affil[11]{Laboratory for Atmospheric and Space Physics, University of Colorado Boulder}
\affil[12]{NASA Goddard Space Flight Center}

\affil[*]{corresponding author(s): Sujit Roy (sujit.roy@nasa.gov) }

\affil[$\dag$]{All authors contributed equally to this work}

\begin{abstract}
This paper introduces a high resolution, machine learning-ready heliophysics dataset derived from NASA's Solar Dynamics Observatory (SDO), specifically designed to advance machine learning (ML) applications in solar physics and space weather forecasting. The dataset includes processed imagery from the Atmospheric Imaging Assembly (AIA) and Helioseismic and Magnetic Imager (HMI), spanning a solar cycle from May 2010 to July 2024. To ensure suitability for ML tasks, the data has been preprocessed, including correction of spacecraft roll angles, orbital adjustments, exposure normalization, and degradation compensation. We also provide auxiliary application benchmark datasets complementing the core SDO dataset. These provide benchmark applications for central heliophysics and space weather tasks such as active region segmentation, active region emergence forecasting, coronal field extrapolation, solar flare prediction, solar EUV spectra prediction, and solar wind speed estimation. By establishing a unified, standardized data collection, this dataset aims to facilitate benchmarking, enhance reproducibility, and accelerate the development of AI-driven models for critical space weather prediction tasks, bridging gaps between solar physics, machine learning, and operational forecasting.

\end{abstract}

\begin{document}

\maketitle

\section{Background \& Summary}

Advancing heliophysics, the study of the Sun and its influence on the solar system, is crucial given space weather's tangible impacts on critical infrastructure like communications, navigation, and power grids~\cite{Schrijver_cohort_2014}. NASA's Solar Dynamics Observatory (SDO) \citep{Pesneletal:2012} continually captures extensive ($\sim$1.5 TB/day), high-quality multi-instrument solar data, turning heliophysics into a data-intensive discipline. This vast observational data from SDO offers a unique opportunity to leverage machine learning (ML) techniques to tackle persistent challenges in solar and heliospheric physics \citep{Asensioetal:2023,roy2024ai}. However, leveraging SDO data presents notable challenges, including specialized preprocessing and domain-aware computational capabilities to homogenize the multi-instrument database \citep{Galvezetal:2019, poduval2023ai}. A publicly available SDO-ML-ready dataset exists \citep{Galvezetal:2019}, but its reduced spatial resolution (512×512) limits the full potential of the original SDO observations.

To address these challenges, we introduce a curated, publicly accessible benchmark dataset from SDO, SuryaBench, comprising high-resolution observations of the solar surface and atmosphere, which can be used to study
diverse solar and heliospheric phenomena such as flares, coronal holes (CH), active regions (AR), sunspots, solar wind, and coronal loops. 
To our knowledge, SuryaBench is the largest curated and homogenized dataset to date, and it preserves the full 4096$\times$4096 native spatial resolution of SDO and provides a consistent 12-minute temporal cadence, enabling high-fidelity analysis for data-driven heliophysics research. 
The dataset is designed to advance data-driven heliophysics research and support operational workflows by offering standardized preprocessing, temporal and spatial homogenization, rich metadata for seamless interoperability, and AI-ready formats, enabling the development and deployment of large-scale machine learning models, specifically self-supervised learning and foundation models \cite{roy2024ai}, and a wide spectrum of heliophysics applications.

SuryaBench is designed to enable the development of advanced, physics-informed models for investigating complex solar phenomena. It features detailed documentation and rich metadata to ensure usability for a wide range of users, including both heliophysics researchers and machine learning practitioners, regardless of their prior domain expertise. To provide a comprehensive and reusable testing environment and facilitate synergistic research, SuryaBench offers standardized application benchmark datasets (hereafter called Datasets, for brevity) for the following six key tasks in heliophysics: (1) solar flare prediction, (2) active region segmentation, (3) active region emergence prediction, (4) coronal magnetic field extrapolation, (5) solar irradiance, and (6) solar wind forecasting. Each application benchmark includes rigorous evaluation protocols and baseline implementations of state-of-the-art machine learning architectures, such as Residual Networks and U-Net. Ultimately, we envision SuryaBench to serve as a robust data resource for diverse, cross-cutting heliophysics tasks with spatio-temporal analysis, multimodal data fusion, and interpretability research. Next, we present a brief overview of selected tasks along with our interdisciplinary motivation for their inclusion. We note that these applications are by no means comprehensive, yet they are relevant to multiple interacting phenomena on the Sun and in the heliosphere, and we envision these to serve as the blueprint for the development of large-scale AI applications using SuryaBench.

At the core of many space weather drivers are active regions (ARs), concentrated areas of magnetic flux that frequently produce solar flares and coronal mass ejections (CMEs), with direct impacts on satellites and terrestrial infrastructure \citep{vanDrielGesztelyiGreen2015}. Within ARs, polarity inversion lines (PILs), which are interfaces where magnetic polarities reverse, are critical sites for energy storage and release, and are strongly associated with solar eruptive activity \citep{schrijver2007pattern,Toriumi2017,Wang2020,Cicogna2021, ji2023systematic}. Twisted and sheared PIL structures can give rise to current sheets and magnetic reconnection, which are central to flare and CME initiation. We provide two tasks related to ARs in Datasets \dsref{1} and \dsref{2} (Sec.~\ref{sec:ar_seg}, \ref{sec:ar_forecast}), where Dataset \dsref{1} is focused on segmentation of ARs with PILs and Dataset~\dsref{2} is focused on AR emergence prediction. That said, the emergence and evolution of ARs also significantly affect coronal dynamics, requiring accurate three-dimensional magnetic field modeling. With Dataset \dsref{3} (Sec.~\ref{sec:3d_extrapolation}), we provide a task on coronal field extrapolation, which supports research on magnetic field extrapolation and AR-induced coronal changes.

Various space weather phenomena have the potential to significantly affect both near-Earth space environments and terrestrial systems. Solar flares, intense eruptions originating in the solar chromosphere and corona, can trigger geomagnetic storms, impacting terrestrial and space-based infrastructure, and posing risks to astronauts \citep{Zhang2001,yasyukevich2018}. Similarly, the solar wind, which is a continuous outflow of charged particles from the solar corona, modulates Earth’s magnetosphere and drives geomagnetic storms with operational implications \citep{cranmer2019properties,Schrijver_cohort_2014,Oughton_dailyimpactspw_2017}. We provide two tasks related to space weather forecasting with Datasets \dsref{4} and \dsref{5} (Sec.~\ref{sec:flares}, \ref{sec:sw_forecast}),  where Dataset~\dsref{4} is focused on flare prediction and Dataset~\dsref{5} is focused on solar wind prediction. Lastly, solar extreme ultraviolet (EUV) irradiance plays a key role in shaping Earth’s ionospheric and thermospheric conditions, influencing satellite drag, communication systems, and GPS accuracy \citep{2005JGRA..110.1312W,2011JGRA..11610309Q,2021JGRA..12628466G,2022FrASS...917103B}. Dataset \dsref{6} (Sec.~\ref{sec:euv_forecast}) addresses EUV nowcasting and forecasting to support satellite operations and mission planning.

\section{Methods}

Our benchmark dataset includes a core imaging data collection, which is primarily designated as input, and six application benchmark datasets, intended as labels, covering different solar physics and space weather applications. In the following subsections, we describe the steps and procedures used to create and curate the data along with the details for each of our application datasets.

\subsection{Core SDO Dataset}

SDO is a NASA Heliophysics flagship mission launched on February 11, 2010 in geosynchronous orbit.  Its main science objectives are to understand how solar magnetism is created, how solar magnetism shapes the extended solar atmosphere that encompasses the entire solar system, and how solar activity affects Space Weather. 

SDO has two imaging instruments: 1.\ The Atmospheric Imaging Assembly (AIA) \citep{Lemenetal:2012}, which measures photometric intensity (per pixel) in the Extreme Ultraviolet (EUV) and UV spectrum.  2.\ The Helioseismic and Magnetic Imager (HMI)\citep{Scherreretal:2012}, which makes spectropolarimetric measurements used to estimate the surface magnetic field (all three components) and the line-of-sight velocity on the solar surface.  Both imagers use 4096$\times$4096 charge-coupled devices (CCDs) to image the solar surface and atmosphere. Table~\ref{tab:SDO} contains details on both AIA and HMI instrumental and data properties.

SDO data is publicly available through the Joint Science Operations Center (JSOC; \url{http://jsoc.stanford.edu}) as time series of various numerical scalar and raster data products. The data series we have used to create our core dataset are \texttt{aia.lev1\_euv\_12s} for EUV channels, \texttt{aia.lev1\_uv\_24s} for a UV channel,  \texttt{hmi.M\_720s} for a line-of-sight (LOS) magnetogram, and \texttt{hmi.B\_720s} for vector magnetograms. 


\begin{table}[tb!]
    \centering
     \caption{Instrumental properties of SDO/AIA and SDO/HMI. Both instruments take 4096$\times$4096 images. AIA measures photometric intensity in EUV for different wavebands.  Its channels, denoted in \AA\,(e.g. 94\AA), indicate the wavelength of peak intensity for each pass-band filter. HMI makes spectropolarimetric measurements around magnetically sensitive spectral emission lines.  We use inversions using these measurements that estimate the three components of the magnetic field (B$_x$, B$_y$, B$_z$), line-of-sight (LOS) magnetic field (B$_{los}$), and LOS velocity (V$_{los}$).  Cadence refers to the time interval between two consecutive images.  We make the distinction between instrumental cadence (12s to 12m) and the dataset cadence (12m).}
    \begin{tabular}{cccccc}\\
        
          Inst.  & Resolution     & Cadence        & Cadence   & Dynamic & Channels\\
            & (photospheric) & (instrument) & (SuryaBench) & range   & \\
        \hline\\
        AIA & 1.2" (725km)   & 12s, 24s  & 12m & $0$ to $16,\!383$ & 94, 131, 171, 193, 211 \\
            &                &      &     &               & 304, 335, 1600 (in \AA) \\
        HMI & 1.0" (870km)   & 45s, 12m  & 12m & $\sim\!\!\pm4,500$ for B  & B$_x$, B$_y$, B$_z$, B$_{los}$ (in G),  \\
        & & & & $\sim\!\!\pm10^4$ for V & v$_{los}$ (in m/s) \\\hline\\
        \end{tabular}
   
    \label{tab:SDO}
\end{table}

\subsubsection{AIA Data Acquisition and Processing}
 
\begin{figure}[!b]
\centering
\includegraphics[width=\textwidth]{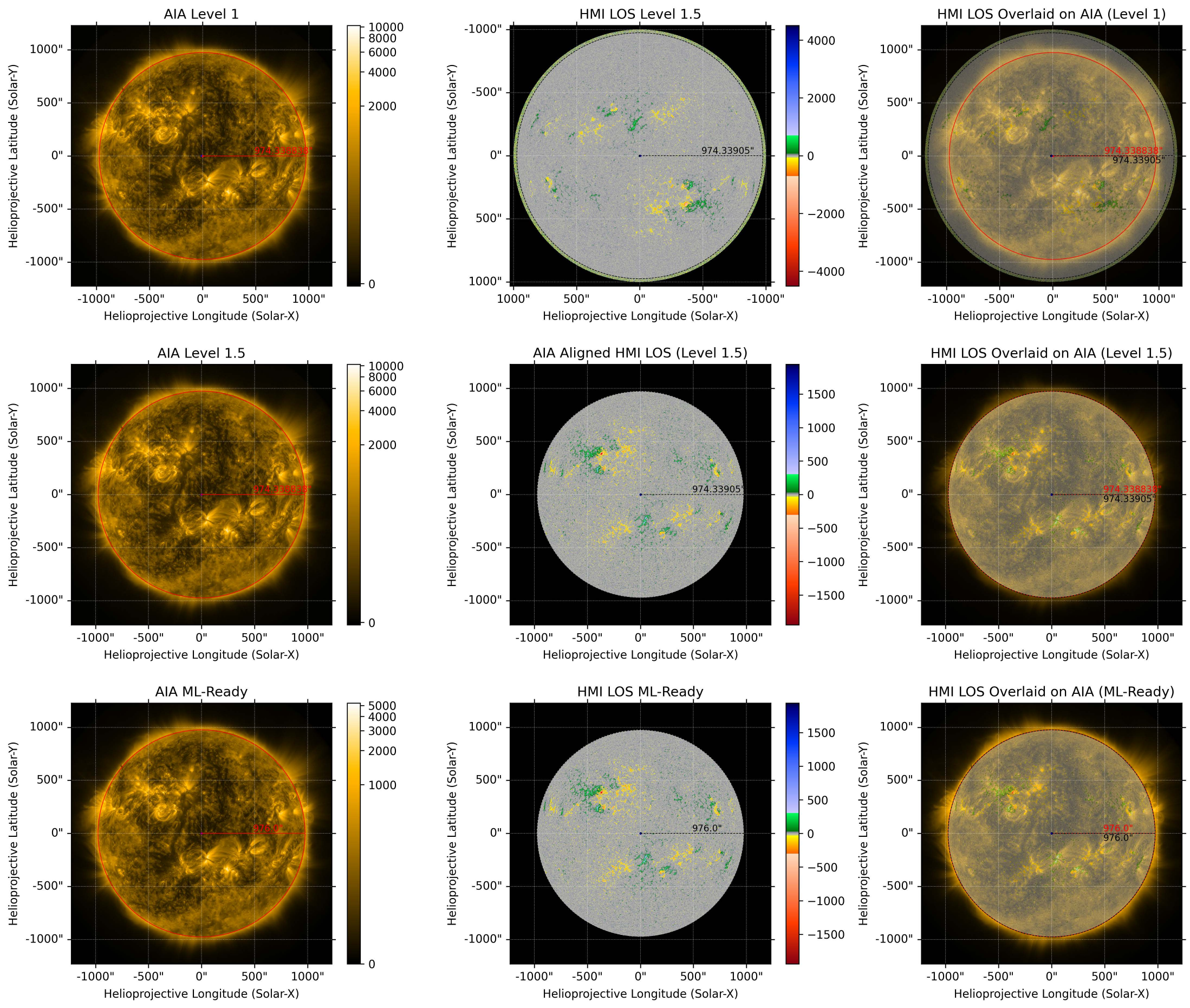} 
\caption{An example of the ML-ready data preparation steps for AIA 171 \AA~ and HMI LOS magnetogram on 2012-01-30 at 22:12 UT. Contours illustrate the image center, solar disk center, disk radius, and solar disk boundary. The top row shows the original AIA Level 1 image, HMI Level 1.5 magnetogram downloaded from JSOC, and HMI overlaid on AIA. The disk centers are misaligned with the image center (unregistered), and one dataset has a 180° roll, with noticeable plate scale differences. The middle row displays the registered AIA Level 1.5 image, HMI aligned with AIA, and HMI overlaid on AIA, showing corrected disk centers and plate scales. The bottom row presents the final ML-ready AIA and HMI images after exposure time normalization and orbital corrections for AIA, with the overlaid image showing proper alignment and a fixed disk radius of 976 arcsecs.}
\label{fig:prepsteps}
\end{figure}

The \texttt{aia.lev1\_euv\_12s} series provided by JSOC contains level-1 data. This means that the images still include the roll angle of the satellite, i.e., the solar north-south axis is not aligned with the vertical y-axis, and each channel may have a slightly different pixel scale. To enhance data accessibility, spatial homogenity, and interoperability, we promoted the AIA data from level-1 to level-1.5 .
The promotion to level-1.5 involves updating the pointing keywords, removing the roll angle, scaling the image to a common pixel scale of 0.6 arcsec per pixel, and translating the image so that the center of the Sun is located in the center of the image. Besides these steps, exposure time normalization is an extra but necessary step during the promotion because AIA measurements have heterogeneous exposure times ranging from 0.05 to 2.9 seconds. We present an example conversion of AIA data from level-1 to level-1.5 in the two top left panels of Figure~\ref{fig:prepsteps} using an  AIA 171$\AA$ image instance. 

\begin{figure*}[tb!]
    \centering
        \includegraphics[width=0.47\textwidth]{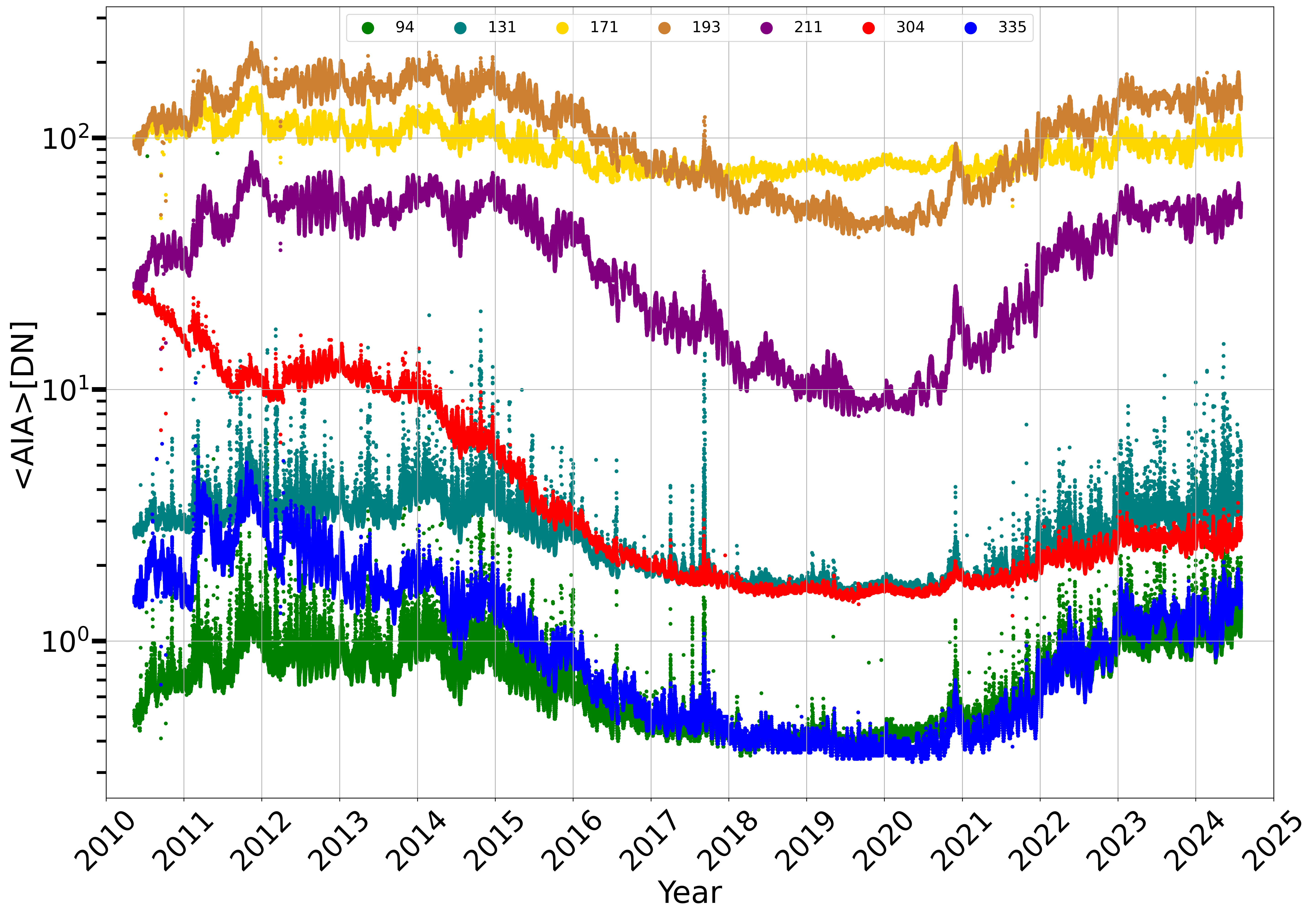}
        \includegraphics[width=0.47\textwidth]{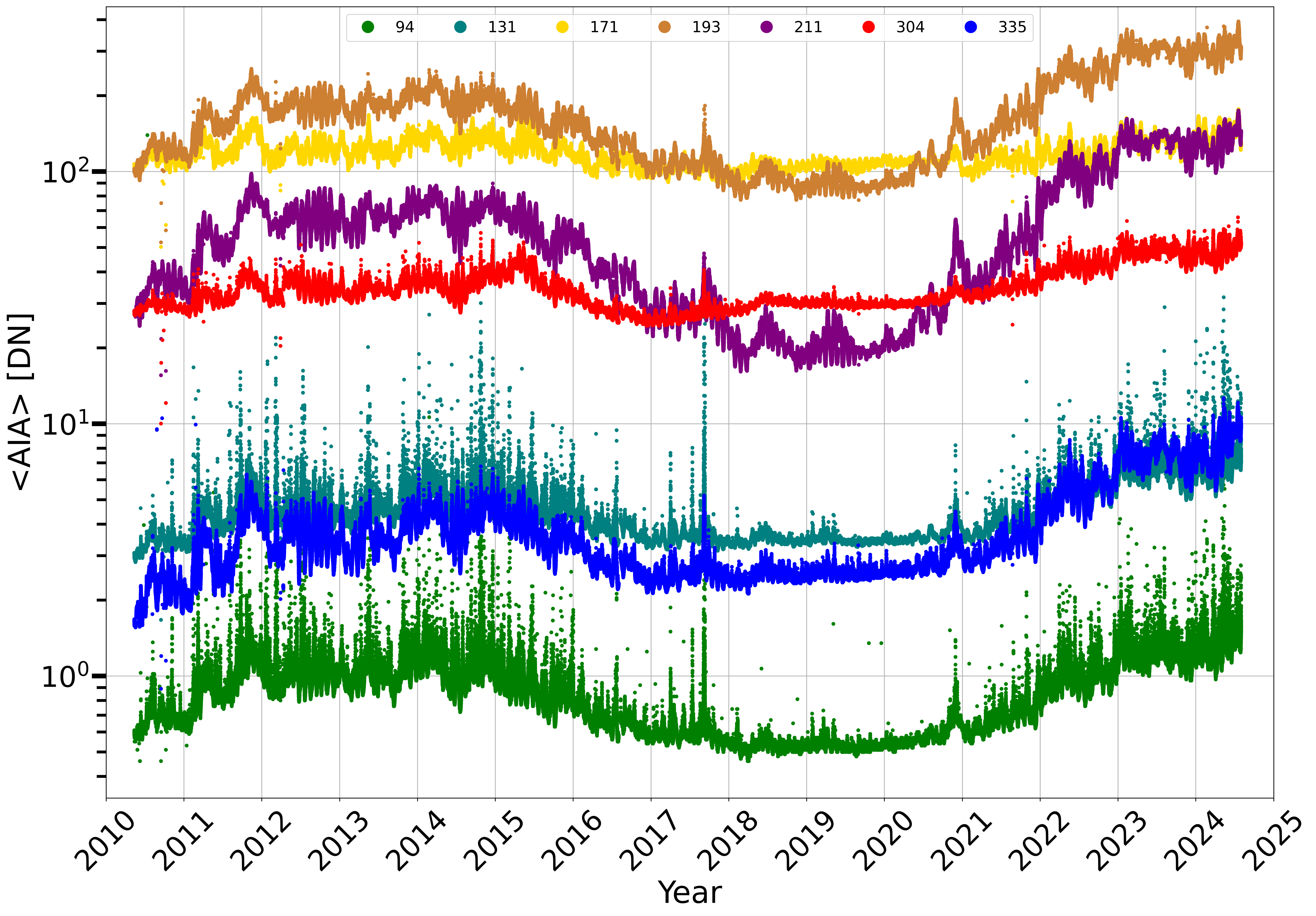}
    \caption{Mean pixel value of full-disk AIA images (extreme ultraviolet channels 94--335\AA, DN/sec) over time before (\textit{left panel}) and after (\textit{right panel}) the degradation correction.}
    \label{fig:degradation_correction}
\end{figure*}

Since the database contains data throughout the SDO lifetime, CCD camera sensor degradation also needs to be taken into account. The table of correction parameters calculated by the AIA science team is made publicly available via JSOC. These parameters are a time series of scalars that can be multiplied by the full disk AIA data to rectify the instrument degradation. However, this method may lead to issues when the corrected values in some of the pixels exceed the instrument saturation value (16,383 for AIA). We post-process and clamp the degradation-corrected image to make sure that none of the pixels reach a value greater than this limit. In Figure~\ref{fig:degradation_correction}, we plot the mean pixel intensity values of the full disk images for each of the seven EUV channels of AIA in level 1 (left panel) and level 1.5 with degradation correction (right panel) data. The variation of mean values over the years also reflects the solar cycle, in which the higher mean values indicate stronger solar activity. We notice that the degradation correction restores the higher activity in solar cycle 25 compared with solar cycle 24.



One final step in making sure that the AIA data is ML-ready is to make the solar disk size the same in the whole database for all wavelengths, correcting for the elliptical orbit of the spacecraft. This step makes the solar disk of fixed radius, 976 arcsecs, in all images. An example of this step is shown in the bottom-left panel of Figure~\ref{fig:prepsteps}.

 \subsubsection{HMI Data Acquisition and Processing}
Though the observable HMI data provided from JSOC are level-1.5, they have a slightly higher resolution of 0.5 arcsec per pixel compared to AIA images.  
Therefore, the HMI data must be re-projected to be spatially aligned with the level-1.5 AIA images, which have 0.6 arcsec per pixel resolution. This re-projection step involves bilinear interpolation when rescaling the high-resolution images to lower-resolution images. Bilinear interpolation estimates the value of a function \( f(x, y) \) at a point \( (x, y) \) that lies between four known grid points \cite{Press_et_al_1992}. In the context of image data, these grid points represent the centers of image pixels, and the function values \( f(x, y) \) correspond to pixel intensities. Let the four surrounding grid points be the corners of a rectangle: the bottom-left \( (x_0, y_0) \), bottom-right \( (x_1, y_0) \), top-left \( (x_0, y_1) \), and top-right \( (x_1, y_1) \), with corresponding function values \( f(x_0, y_0) \), \( f(x_1, y_0) \), \( f(x_0, y_1) \), and \( f(x_1, y_1) \), respectively. If we consider a desired point \( (x, y) \) within this rectangle, such that \( x_0 \leq x \leq x_1 \) and \( y_0 \leq y \leq y_1 \), the interpolated value is given by:
\[
f(x, y) = (1 - t)(1 - u)\,f(x_0, y_0) + t(1 - u)\,f(x_1, y_0) + (1 - t)u\,f(x_0, y_1) + tu\,f(x_1, y_1).
\]
where
\[
t = \frac{x - x_0}{x_1 - x_0}, \quad u = \frac{y - y_0}{y_1 - y_0}
\]
represent the normalized distances of the point \( (x, y) \) along the \( x \)- and \( y \)-axes, respectively. The \textit{reproject\_to} function available in SunPy \cite{sunpy_community2020} was used for this step.
An example of the re-projection is shown in the top two panels in the middle column of Figure~\ref{fig:prepsteps}. 
Finally, similarly to AIA images, the HMI images also needed to be corrected for elliptical orbit variation and fixed to a solar disk size of 976 arcsecs throughout the database. This step is shown with an example in the bottom panel in the middle column of Figure~\ref{fig:prepsteps}. 
The panels in the right column of this image show that after our processing, the solar disk in the AIA and HMI images are well aligned.

\subsubsection{Temporal Alignment of HMI and AIA Data}
In our database, we have decided to keep the temporal resolution of the time series at 12 minutes. This is because the low-noise LOS and vector magnetograms are available with this cadence from HMI. 
For each of the timestamps in \texttt{hmi.M\_720s} series, we find the corresponding AIA data at that time and add it to our database. If the quality of the AIA data for any of the eight wavelengths is not good at an exact timestamp of HMI, we search for a time within two minutes of the timestamp that contains good quality AIA data in all eight EUV channels and include that data in our database. If we are unable to find good-quality AIA data within the two-minute range, we do not include the data in our database. We consider the quality of the AIA data to be good if the QUALITY flag in the file header is equal to zero. Non-zero values for the QUALITY flag indicate different operational items (e.g., off-pointing or defocusing of the satellite) lowering the quality or significant missing data. 
In Figure~\ref{fig:baddata} (see the Supplementary Information), we give examples of when the AIA data was found to be of bad quality.

\subsection{Application Benchmark Datasets}

In this section, we will describe our application benchmark datasets in detail. As shown in Table \ref{tab:downstream}, multiple space weather and heliophysics applications are broadly oriented towards understanding the deposition of energy in the solar surface and atmosphere (\dsref{1} \& \dsref{2}; Sec.~\ref{sec:ar_seg} \& \ref{sec:ar_forecast}), how solar magnetism structures the solar atmosphere (\dsref{3}; Sec. \ref{sec:3d_extrapolation}), how energy is released in the form of space weather (\dsref{4}; Sec.~\ref{sec:flares}), and how space weather affects the solar magnetosphere (\dsref{5};  Sec.~\ref{sec:sw_forecast}) and ionosphere (\dsref{6}; Sec.~\ref{sec:euv_forecast}). We note that we created baseline models and evaluated them. Our model architectures, formulated problems, evaluation measures, and results can be found in the Supplementary Information document.

\subsubsection{Active region segmentation}\label{sec:ar_seg}

\begin{figure}[tb!]
    \centering
    \includegraphics[width=\linewidth]{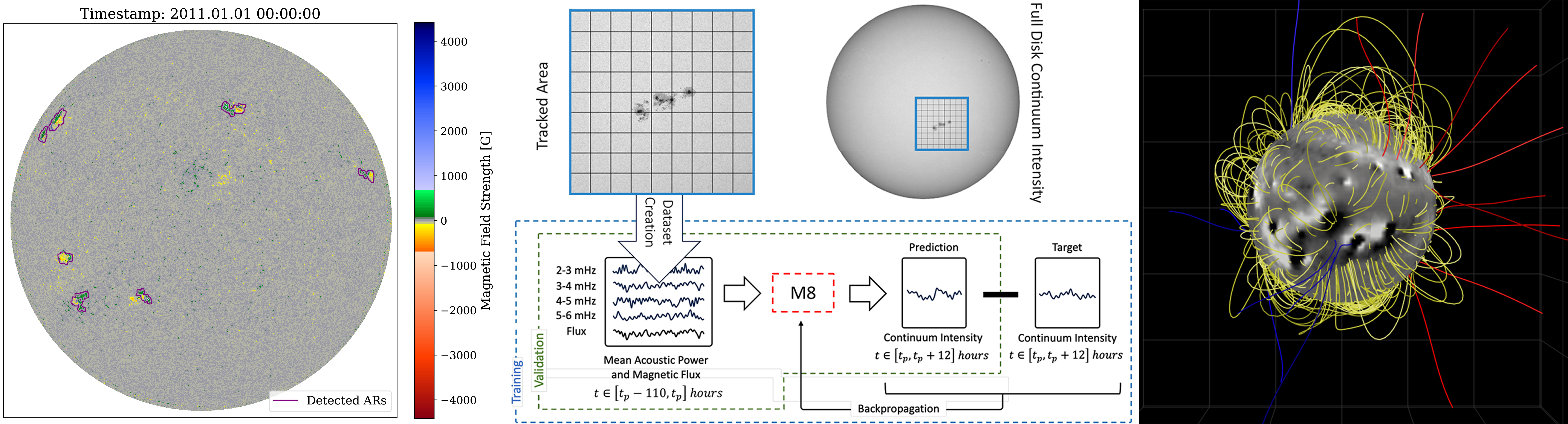}
    \caption{\textbf{\textit{Left: }} Example of an AR mask for segmentation. The magnetic field strength from the line-of-sight magnetogram on 2011-01-01 00:00:00 is shown using a blue-to-red color scale. Detected active regions, which have magnetic flux greater than 50 Gauss or less than –50 Gauss and contain polarity inversion lines, are highlighted with purple contours. \textbf{\textit{Middle:}} Example of AR intensity time-series for emergence forecast. Tracked regions are split into smaller tiles, and the timeline datasets were created by averaging the values of each tile. The timelines are used as inputs during the training and validation/testing. \textbf{\textit{Right:}} Example coronal magnetic field extrapolated from measurements using WSA. The yellow magnetic field lines indicate closed field lines that loop back to the surface of the sun, which the red/blue lines are positive/negative polarity open field lines which extend past the source surface into interplanetary space.}
    \label{fig:AR_3D}
\end{figure}

Active regions (ARs) are often identified using intensity thresholding on magnetograms, white-light or EUV images \cite{turmon2010automated, verbeeck2014spoca, Caballero2013}. 
 In this work, ARs containing polarity inversion lines (PILs) are identified using full-disk line-of-sight (LoS) magnetograms from the Solar Dynamics Observatory (SDO). 
Our previous PIL detection method by \cite{cai2020framework, ji2023systematic} relies on AR patches. Extending this, our approach uses full-disk LoS magnetogram rasters with a resolution of 4096×4096. We first generate two binary maps, corresponding to the positive and negative polarity regions, by applying a magnetic field strength threshold of $+50$ and $-50$\,Gauss, respectively. To remove small, noisy patches in positive and negative polarity region maps, we apply a size filter that excludes regions smaller than 100 pixels (approximately $13.3\ \mathrm{Mm}^2$ of photospheric area). Next, we dilate the binary images using a rectangular filter of size 10 pixels. Finally, we identify the intersection of the dilated positive and negative polarity regions, which corresponds to areas containing PILs. Only ARs that include PILs are reported as regions of interest. The eventual AR masks are 2D bitmaps (containing zeros and ones), representing the locations of active regions with PILs, and have a size of 4096$\times$4096. In other words, ARs without the presence of a strong PILs will be omitted from the full-disk mask. In Figure \ref{fig:AR_3D}-left, the binary mask is overlaid on the original line-of-sight magnetogram from 2011-01-01 00:00:00. The regions outlined in purple indicate the detected active regions.

The ARPIL dataset covers between January 2011 to December 2024. The detection method is applied to each valid LoS magnetogram hourly. In total, we have 119,454 full-disk AR binary masks covering 14 years, each of them with a size of 4096$\times$4096.

\unskip

\subsubsection{Active region emergence forecast}\label{sec:ar_forecast}

We select 50 ARs that appear on the solar surface within 30 degrees longitude from the central meridian between March 1st, 2010 and June 1st, 2023, persisted for more than 4 days, and reached a total area of 200 millionths of the solar hemisphere. For each one of these 50 ARs, the same five-step pipeline is followed: (1) tracking areas of 512 by 512 pixels of the SDO/HMI magnetic flux, Doppler velocity, and continuum intensity, (2) creating acoustic power maps from the Doppler velocity, (3) downsampling the data to timelines by splitting the tracked region in a 9 by 9 grid, (4) removing the solar sphere geometric effects, and lastly (5) calculating the time evolution of continuum intensity.

Before we create the acoustic maps, we use the dopplergrams series, representing the frames throughout the life of the AR on the solar disk, and create a difference series by subtracting consecutive frames.  By working with dopplergram differences, we remove the background solar rotational signal:
\vspace{-10pt}

\begin{equation}
\Delta V_{\text{dop}}[i,x,y] = V_{\text{dop}}[i+1,x,y] - V_{\text{dop}}[i,x,y], \quad \text{for } i = 1, \ldots, 639, \quad (x,y) \in [1,512]^2.
\end{equation}

Each element of $\Delta V_{\text{dop}}$ represents the difference between consecutive dopplergrams at each pixel location $(x, y)$. Subsequently, for each pixel, we calculate the Fourier power spectrum of the time-series data in $\Delta V_{\text{dop}}[:,x,y]$. Let $dt = 45$ sec (the sampling interval), $T = 28800$ sec (each .fits file is tracks the active region for 8 hours), and $\mathcal{F}$ denote the real-valued one-sided Fourier transform (\textit{np.fft.rfft} in Python):

\vspace{-10pt}

\begin{equation}
V_{\text{dop}}^{\text{FFT}}[k,x,y] = \left(\frac{dt^2}{T}\right) \left| \mathcal{F} \{ \Delta V_{\text{dop}}[:,x,y] \} [k] \right|^2, \quad \text{for } k = 1, \ldots, 320, \quad (x,y) \in [1,512]^2.
\end{equation}


We calculate the Fourier power spectrum of the time series data. For every timeframe, we calculate the integral along the frequency axis to construct a power map that represents the spectral power in the chosen frequency range at each spatial location on the solar disk. Given a power map series, which represents the temporal evolution of the spatial power distribution for a particular frequency range (e.g., $2-3 mHz$) over the solar disk, we divide the solar disk into a grid of smaller, equally-sized tiles.  For each of these tiles, we extract the corresponding pixels from all frames of the power map series.

Subsequently, we calculate the mean power within each tile for each frame. The temporal mean power within a tile is calculated by taking the average power over all pixels within the tile for each frame.  This procedure results in a one-dimensional time series per pixel, representing the temporal evolution of power within each segment on the solar disk. In parallel, we calculate the total continuum intensity, as well as total magnetic flux for each tile to produce labeled acoustic power time series and magnetic and continuum intensities.  The combination of the acoustic power timeseries and continuum intensity time series (as shown in Fig.~\ref{fig:AR_3D}-middle) form the core of this dataset.

Threfore, this dataset includes timeseries for 50 emerging ARs. Each has 6 channels (4 acoustic power channels, magnetic flux, and continuum intensity). Each time series has 240 timestamps.  The dynamic range goes between $-7.5\times10^{7}$ and $5.8\times10^{7}$ for the acoustic power channels, $-1.4\times10^{2}$ to $5.3\times10^{2}$ for magnetic flux, and $-1.7\times10^{4}$ and $4.0\times10^{3}$ for continuum intensity. More information about the AR emergence dataset can be found in \cite{kasapis2023predicting}.

\unskip

\subsubsection{Coronal field extrapolation}\label{sec:3d_extrapolation}

To model the 3D structure of the coronal magnetic field, it is necessary to first estimate a full coverage (180$^{\circ}$ latitude, 360$^{\circ}$ longitude), and subsequently model the transformation of a solar surface boundary condition into a magnetic field that extends into the atmosphere. To do this, we use a coupled simulation known as ADAPT-WSA.

For the first task, we use the Air Force Data Assimilative Photospheric Flux Transport (ADAPT; \cite{arge2010air,arge2011improving,arge2013modeling};  \cite{hickmann2015data}), which is a data assimilation model that uses near-side photospheric magnetic field measurements from such instruments as HMI and globally solves a system of magnetic flux transport equations \cite{worden2000evolving}. These equations describe the time-dependent evolution of the photospheric magnetic field, including such effects as differential rotation, and meridional and supergranular flows. ADAPT is an ensemble model which generates 12 realizations (variations) per timestep. Each realization represents processing using different values of unobservable subsurface physics in the ADAPT simulation, which is itself an ensemble Kalman filter.

For the second task, we use the Wang--Sheeley--Arge (WSA; \cite{arge2000improvement,arge2003improved,mcgregor2008analysis}) model, which  makes it possible to calculate the global coronal field using a coupled Potential Field Source Surface and Potential Field Current Sheet (PFCS) approaches \cite{schatten1969model,wang1992potential}. In practice, the WSA model solves the equations $\nabla \times \mathbf{B} = \mathbf{0}$  and $\nabla \cdot \mathbf{B} = 0$, where $\mathbf{B}$ is the coronal magnetic field, up to a spherical boundary known as "the source surface" and where the field becomes radial, set here at 2.51 $R_\odot$ (solar radii). The coronal field extrapolations are themselves encoded using the spherical harmonics.



This dataset includes an 11-year (full solar cycle) span of ADAPT-WSA runs powered by the HMI magnetogram data at daily cadence. 
Daily cadence is chosen to provide a sufficient diversity of magnetic topologies in the training dataset and avoid nearly identical training samples. The ADAPT-WSA runs are split into an ensemble composed of 12 realizations per the ADAPT ensemble. The potential field solution at each timestep is encoded in signed spherical harmonic coefficients, normalized using the Schmidt formulation and truncated after the 90th order.  An example magnetic field solution is displayed in Fig.~\ref{fig:AR_3D}-right,  where the plotted field lines were traced using the spherical harmonics to evaluate $\mathbf{B}$ at each position.

This dataset includes 51,156 sets of harmonic coefficients, each has 2 channels (G and H coefficients), and contains 4,186 harmonic coefficients.  The dynamic range goes between $-4.3x10^{3}$ to $4.3x10^{3}$.\unskip

\unskip

\subsubsection{Flare forecasting}\label{sec:flares}

Although the volume of observational data has significantly increased, accurate operational prediction solar flares remains a challenging task.  Solar flares are monitored by the Geostationary Operational Environmental Satellites (GOES), measuring the X-ray intensity emitted by the Sun. The National Oceanic and Atmospheric Administration (NOAA) classifies solar flares logarithmically into five major classes --A, B, C, M, and X, based on their peak X-ray intensity in the 1–8$\text{\AA}$ wavelength range \citep{fletcher2011observational}. The strength of a flare within a class is indicated by a numerical suffix ranging from 1.0 to 9.9, which represents the factor by which the event is stronger than the base intensity in that class (e.g., M5.2 is 5.2 times as strong as M1.0).  Flares above C-class, particularly M- and X-class flares, are of primary interest due to their significant terrestrial impact, yet the scarcity of stronger events pose a substantial class imbalance challenge. 

\begin{figure}[bt!]
    \centering
    \includegraphics[width=\linewidth]{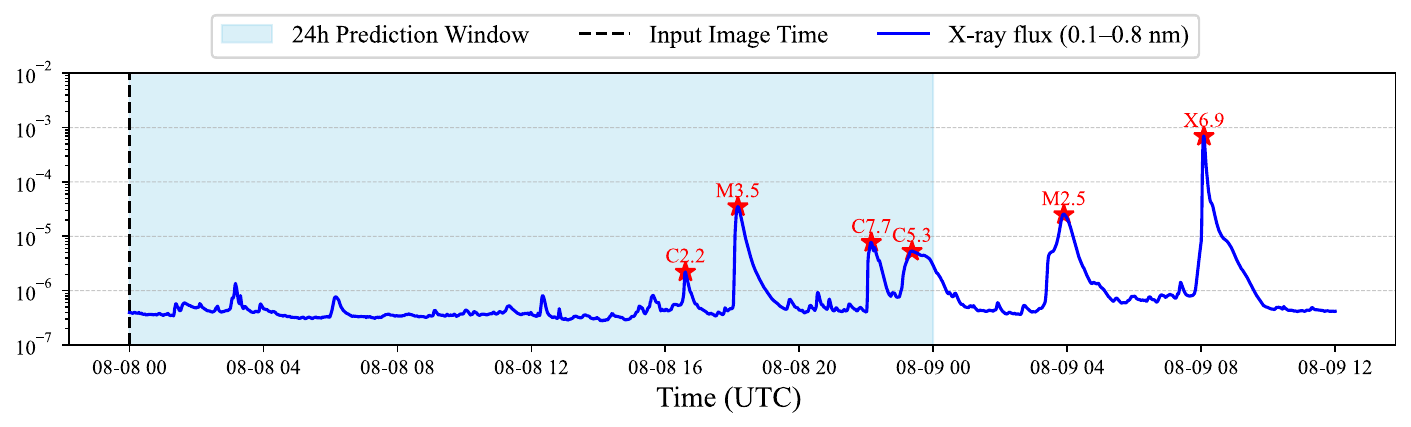}
    \includegraphics[width=0.59\linewidth]{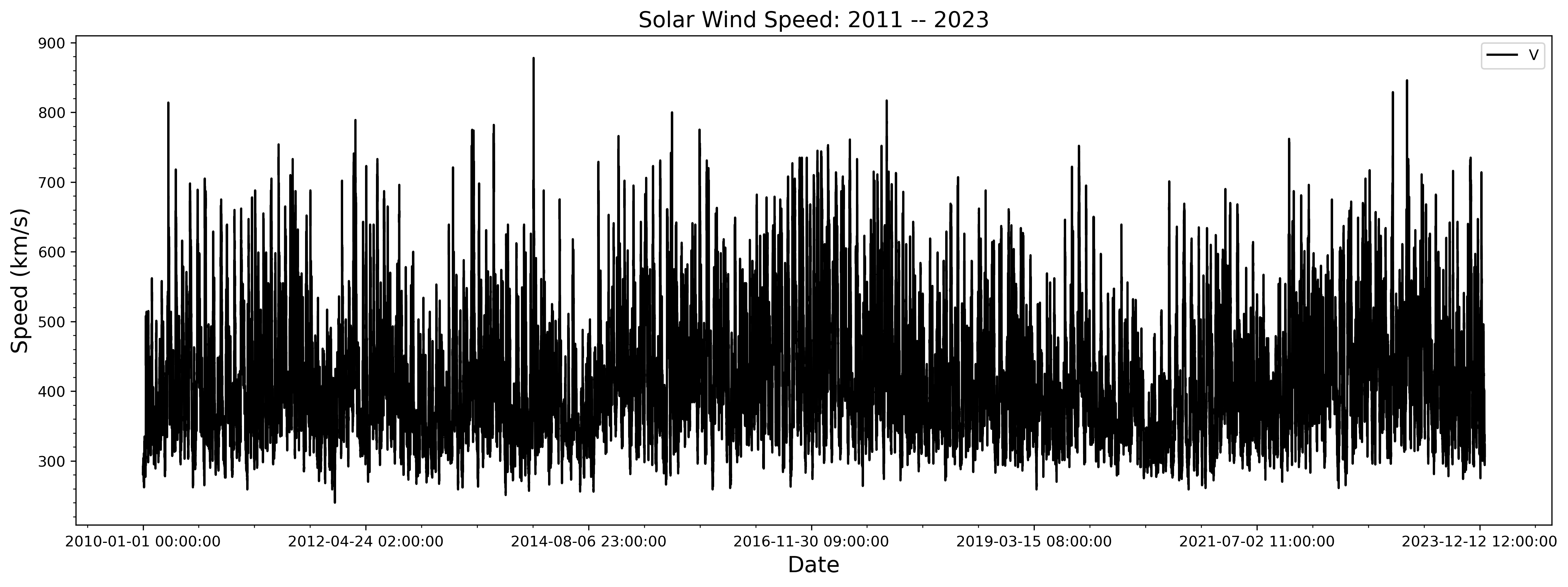}
    \includegraphics[width=0.35\linewidth]{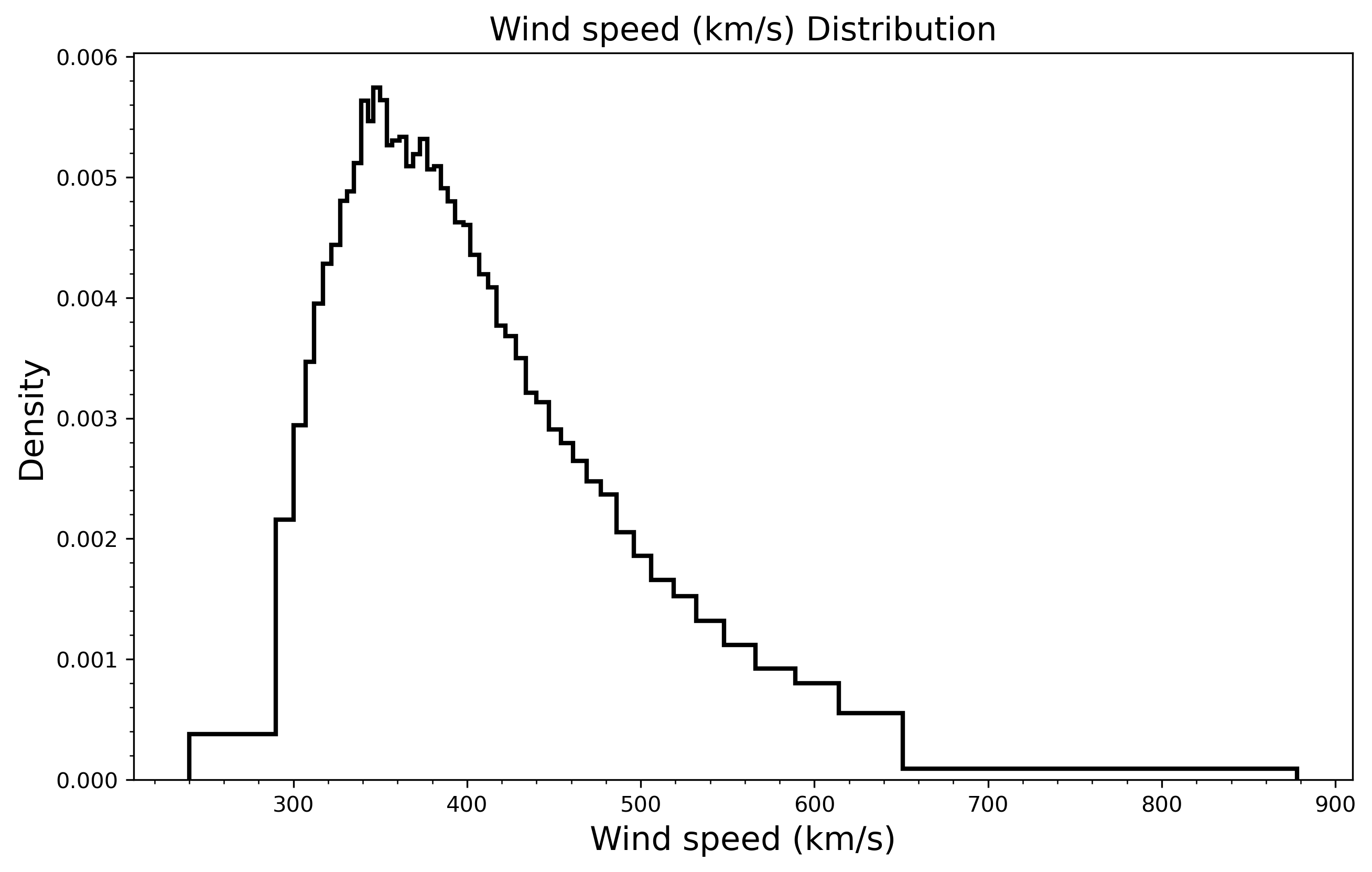}
    \caption{\textbf{\textit{Top:}} Labeling inputs with the flare index. The 24-hour prediction window may include multiple flares. Two types of flare indices are defined: the maximum flare intensity and the cumulative flare intensity. Inputs are sampled at an hourly cadence, and the prediction window shifts accordingly based on the input time.  \textbf{\textit{Bottom-Left:}} The solar wind speed measurements at L1 as a time series. \textbf{\textit{Bottom-Right:}} Distribution of solar wind speed measurements.}
    \label{fig:flare-wind}
\end{figure}

In this dataset, the input instance at time $t_i$ is associated with a prediction window spanning from $t_i$ to $t_i+24h$. Each window may contain zero or more solar flares. Note that we use the start time of the flares to determine if they are within a prediction window. Only flares greater than C-class are considered in this application due to the under-reporting of lower intensity flares. 
Each input, sampled at an hourly cadence, is labeled in two ways: (1) by the \textit{maximum flare intensity}, as defined in Eq.~\hyperref[eq:max-label]{2}, and (2) by the \textit{cumulative flare intensity}, as defined in Eq.~\hyperref[eq:cum-label]{3}. Maximum flare intensity is the label corresponding to the flare with the highest intensity occurring within the prediction window. 
\begin{equation} \label{eq:max-label}
    L_{max}(t_i) = class(\max_{f_j \in F_{t_i}^{C^+}} \mathbf{pxf}(f_j))
\end{equation}
, where $class(\cdot)$ returns the GOES class corresponding to the peak X-ray flux for the maximum intensity flare.
In Figure \ref{fig:flare-wind}-top, we demonstrate an example prediction window covering four flares shown. This instance is labeled as `M3.5', which is the flare with the maximum intensity.  

Cumulative flare intensity considers the cumulative effect of all the $\geq$C-class flares in the prediction window. As mentioned earlier, flare sub-classes (e.g., C\textbf{5.2}, M\textbf{1.0}) are indicated by a numerical suffix ranging between 1.0 and 9.9. To create the the cumulative intensity label, we get the weighted sum of these numerical suffixes/subclass values, as described in Eqs. \hyperref[eq:class-weight]{3} and \hyperref[eq:cum-label]{4}. 
 
\begin{equation}
    \label{eq:class-weight}
    S(f_j) =
    \left\{
    \begin{array}{ll}
    0 & \text{if } \mathbf{pxf}(f_j) < 10^{-6} \\
    1 & \text{if } 10^{-6} < \mathbf{pxf}(f_j) \leq 10^{-5} \\
    10 & \text{if } 10^{-5} < \mathbf{pxf}(f_j) \leq 10^{-4} \\
    100 & \text{if } \mathbf{pxf}(f_j) > 10^{-4}
    \end{array}
    \right. ~,
\end{equation}
where $S(f_j)$ returns the weight for the flare event $f_j$. In other words, to differentiate the contribution of C-, M-, and X-class flares, weights of 1, 10, and 100 are applied to their respective subclass values. The weighted sum $L_{cum}(t)$ is then calculated as:

\begin{equation} \label{eq:cum-label}
    L_{cum}(t_i) = \sum_{f_j \in \mathcal{F}_{t_i}^{C^+}} S(f_j) \cdot v(f_j),
\end{equation}
where $v(f)$ denotes the subclass value for $f_j$. For example, in Figure~\ref{fig:flare-wind}, four subclass values from four flares within the prediction window (with blue background) are considered. The cumulative flare intensity is 50.2, calculated as $2.2\ (\text{C2.2}) + 10 \times 3.5\ (\text{M3.5}) + 7.7\ (\text{C7.7}) + 5.3\ (\text{C5.3})$.

For creating labels for binary classification, we use two thresholds for $L_{max}$ and $L_{cum}$ corresponding to the equivalent strength of an M1.0-class flare. In other words, we create two binary labels checking (1) $L_{max} > M1.0$ and $L_{cum} > 10$. The flare forecasting labels span from May 2010 to December 2024. There are total 128,352 labels in the dataset.

\unskip

\subsubsection{Solar wind forecasting}\label{sec:sw_forecast}

The solar wind is a stream of charged particles that emanates from the Sun. The interaction of solar wind with Earth's magnetic field results in the formation of near-Earth space weather. Solar wind interactions are known to drive geomagnetic storms, wherein the Earth's magnetic field is perturbed, inducing electrical currents that affect satellites, power grids, oil pipelines, etc., and potentially resulting in economic and livelihood impacts~\cite{Schrijver_cohort_2014,Oughton_dailyimpactspw_2017}. The solar wind is known to have origins from expansive regions very low in the solar atmosphere, undergoing acceleration as it propagates outwards through the chromosphere, transition region, and corona \cite{cranmer2019properties}, and then becoming free streaming in the interplanetary medium. In a more global sense, the solar wind shows statistical associations with morphological structures in the solar atmosphere.

For this application benchmark dataset, we use space-based particle data, measured by the Advanced Composition Explorer (ACE; \cite{ace}), which are made available through the NASA OMNIWEB database~\footnote{\url{https://omniweb.gsfc.nasa.gov/}}. The OMNI data consist of solar wind speed measurements at the L1 point of the Sun-Earth system in space, and are time-shifted to be at the nose of the Earth's bow shock. Hence, we essentially have a scalar measurement across all time, resulting in a vector of measurements. These data spans 14 years from 2010 to 2023, at a time cadence of 1 hour. The time series of the full dataset is shown in Fig.~\ref{fig:flare-wind}-bottom-left.
The solar wind speed, as seen in Figure~\ref{fig:flare-wind}-bottom-right, shows a form of long tailed distribution, with a peak at $\approx400$ km/s. This dataset includes 120,748 measurements of the solar wind speed. The dynamic range goes between $2.4\times10^{2}$ to $8.8\times10^{2}$ km/s.

\unskip

\subsubsection{Solar EUV spectra modeling}\label{sec:euv_forecast}

The Extreme Ultraviolet Variability Experiment (EVE; \cite{2012SoPh..275..115W}) aboard NASA's Solar Dynamics Observatory (SDO) was developed to provide high-resolution, high-cadence measurements of full-disk solar EUV irradiance. In this dataset, we focus on the high energy part of the EUV, captured by a submodule of EVE called MEGS-A, which has a 10-second cadence \citep{EVEL2Bv8README}. We apply metadata-based screening using the quality flags provided in the EVE Level 2 data products \citep{EVEL2Bv8README}. The \verb|SC_FLAGS| byte identifies potential obstructions or pointing issues during observation. Only spectra flagged as \verb|0| (clear, unobstructed) are retained. We compute 1-minute averaged EVE spectra to reduce noise and facilitate temporal alignment with AIA image data. We then match the 12 minute cadence of the SDO/AIA image cubes.  Timestamps in which AIA frames affected by saturation (common during large flares), diffraction patterns, and instrument exposure anomalies are flagged and excluded. The resulting dataset contains hundreds of thousands of temporally aligned AIA image cubes and EUV spectra, covering Solar Cycle 24 and parts of Solar Cycle 25, and includes both quiet-Sun and active-region conditions.  All irradiance values are first corrected to 1 astronomical unit (AU) to remove the influence of Earth–Sun distance variations.

Event selection and temporal stratification were performed to construct a balanced and representative training set spanning a wide range of solar conditions. For active periods, we rely on the GOES X-ray Sensor (XRS; \cite{2009SPIE.7438E..02C}) flare catalog to identify flaring events. These events are binned according to the integrated soft X-ray flux and stratified into percentiles, rather than raw flare class labels, to ensure a more uniform representation of flare energetics and avoid over-representation of weaker, more frequent flares. For quiet Sun conditions, where changes in irradiance are dominated by solar rotation and large-scale structural evolution, data are sampled at regular 1-day intervals to capture the modulations introduced by active region transit across the solar disk. This dual strategy ensures adequate exposure to both high-energy transient events and slowly varying background structures.

This dataset includes 189,397 EVE spectra, each with 1343 spectral channels.  The dynamic range goes between $1.0\times10^{-9}$ to $1.1\times10^{-2}$.

\unskip

\subsection{Summary}

SuryaBench captures diverse solar phenomena across a full solar cycle, with high-resolution multi-instrument observations and rigorous standardization.
By integrating data from AIA and HMI
with consistent preprocessing into a unified ML-ready format, the dataset provides a high-fidelity view of solar activity. This uniformity enhances data quality and reproducibility, enabling cross-comparison of events (flares, active region evolution, coronal dynamics) and cultivating deeper insight into the dynamics of solar activity and space weather. We believe that the inclusion of curated benchmarks and baseline model results for tasks such as solar flare prediction, coronal field extrapolation, and active region segmentation underscores SuryaBench’s value to the machine learning community, and we envision that these application benchmarks will establish clear performance baselines and spurring the development of advanced models. The dataset breadth and ML-focused design bridge the heliophysics and AI, accelerating progress in space weather predictive modeling.

\section{Data Records}

\label{sec:Stats}

The SuryaBench datasets contain ML-ready heliophysics data captured from May 13, 2010, to December 31, 2024, with a 12-minute cadence. The datasets (both core and application benchmark datasets) are publicly available on Huggingface as a data collection \href{https://huggingface.co/collections/nasa-ibm-ai4science/suryabench-689276db205c56e81c10070c}{Suryabench}.
During this collection interval, there are about 6\% data is missing due to either unavailability or poor quality. The processed level-1.5 AIA and HMI data are stored in hourly netCDF files in \texttt{float32} format, with data shape of [13, 4096, 4096]. Each netCDF file is about 600 MB, and the total size of the data for training is approximately 360 TB. We have divided the data into training (2010-2018), validation (2019), and test (2020) sets, which include 379,920, 43,680, and 43,800 files, respectively.

\begin{table}[!bt]
\centering
\caption{Statistic summary of auxiliary datasets. The shape of all datasets has been cast with the following dimensions: N $\rightarrow$ Number of datapoints, C $\rightarrow$ Number of channels, T $\rightarrow$ Number of timestamps, H $\rightarrow$ height, and W $\rightarrow$ width.}
\setlength{\tabcolsep}{4pt}
\renewcommand{\arraystretch}{1.15}
\resizebox{\linewidth}{!}{%
\begin{tabular}{|c|p{0.20\textwidth}|p{0.28\textwidth}|c|c|c|c|c|p{0.22\textwidth}|}
\hline
\textbf{} & \textbf{Application} & \textbf{Broader Topic} & \textbf{N} & \textbf{C} & \textbf{T} & \textbf{H} & \textbf{W} & \textbf{Dynamic Range} \\ \hline
\dslabel{1} &
\begin{tabular}[c]{@{}l@{}}AR\\segmentation\end{tabular} &
\begin{tabular}[c]{@{}l@{}}Magnetic energy\\in solar atmosphere\end{tabular} &
109{,}175 & 1 & 1 & 4{,}096 & 4{,}096 & 0,1 \\ \hline
\dslabel{2} &
\begin{tabular}[c]{@{}l@{}}AR emergence\\forecasting\end{tabular} &
\begin{tabular}[c]{@{}l@{}}Magnetic energy\\in solar atmosphere\end{tabular} &
50 & 6 & 240 & 1 & 1 &
\begin{tabular}[c]{@{}l@{}}$-1.7\times10^{4}$\\to $4.0\times10^{3}$\end{tabular} \\ \hline
\dslabel{3} &
\begin{tabular}[c]{@{}l@{}}Coronal field\\extrapolation\end{tabular} &
\begin{tabular}[c]{@{}l@{}}Coronal structure\\\& magnetism\end{tabular} &
51{,}156 & 2 & 1 & 4{,}186 & 1 &
\begin{tabular}[c]{@{}l@{}}$-4.3\times10^{3}$\\to $4.3\times10^{3}$\end{tabular} \\ \hline
\dslabel{4} &
\begin{tabular}[c]{@{}l@{}}Flare\\forecasting\end{tabular} &
\begin{tabular}[c]{@{}l@{}}Space weather\\forecasting\end{tabular} &
128{,}352 & 1 & 1 & 1 & 1 & 0,1 \\ \hline
\dslabel{5} &
\begin{tabular}[c]{@{}l@{}}Solar wind\\forecasting\end{tabular} &
\begin{tabular}[c]{@{}l@{}}Solar forcing\\of magnetosphere\end{tabular} &
120{,}748 & 1 & 1 & 1 & 1 &
\begin{tabular}[c]{@{}l@{}}$2.4\times10^{2}$\\to $8.8\times10^{2}$\end{tabular} \\ \hline
\dslabel{6} &
\begin{tabular}[c]{@{}l@{}}EUV\\forecasting\end{tabular} &
\begin{tabular}[c]{@{}l@{}}Solar forcing\\of ionosphere\end{tabular} &
189{,}397 & 1{,}343 & 1 & 1 & 1 &
\begin{tabular}[c]{@{}l@{}}$1.0\times10^{-9}$\\to $1.1\times10^{-2}$\end{tabular} \\ \hline
\end{tabular}%
}
\label{tab:downstream}
\end{table}

\section{Technical Validation}

We used the SuryaBench datasets to validate against state-of-the-art models commonly used by the machine learning and heliophysics communities. All experiments were performed on 4 Nvidia A100 GPUs with 80 GB of memory. This evaluation helps establish reference points for future research by comparing performance across widely adopted architectures. To create the baseline on the SDO dataset, we framed it as a forecasting problem. Given two input tasks we predicted the next time step. By training on 4 years' worth of data, the modified long-short Spectral Transformer \cite{roy2024ai} model demonstrated strong performance after training for just 20 epochs on four years of data. For the AIA bands, the model achieved high structural similarity index (SSIM) scores of 0.83 for band 171\AA, 0.90 for band 193\AA, and 0.86 for band 211\AA, while the remaining bands showed SSIM values ranging from 0.4 to 0.65. The corresponding root mean squared error (RMSE) values were 0.11 for band 171\AA, 0.095 for band 193\AA, and 0.10 for band 211\AA. When applied to the HMI channel, the model achieved an SSIM of 0.73 and an RMSE of 0.65. These results indicate the model’s ability to accurately reproduce both large-scale and fine-scale solar features, including active regions with higher magnetic field strengths.

The tables below summarize baseline results for two core tasks: solar wind forecasting and binary solar flare prediction. Solar wind forecasting performance is reported using RMSE, MAE, and validation loss for ResNet and U-Net based encoder-decoder models. For flare prediction (classification task), we evaluate models, including AlexNet~\cite{krizhevsky2012imagenet}, MobileNet~\cite{howard2017mobilenets}, and ResNet~\cite{he2016deep} variants, using popular forecast skill scores True Skill Statistic (TSS), Heidke Skill Score (HSS), Composite Skill Score (CSS), and F1-macro, similar to \cite{Pandey2024ecml}. These baselines provide a standardized performance floor for advancing heliophysics AI.

\begin{table}[H]
\centering
\caption{{Baseline performance for (a) solar wind prediction and (b) solar flare classification using common deep learning models on test data}}
\vspace{0.5em}
\begin{minipage}{0.48\textwidth}
\centering
\begin{tabular}{lccc}
\toprule
\textbf{Model} & \textbf{RMSE} & \textbf{MAE} & \textbf{Val Loss} \\
\midrule
UNet           & 0.1499  & 0.1116  & 0.0225 \\
AttentionUNet  & 0.1449  & 0.1157  & 0.0225 \\
ResNet18       & 0.2108  & 0.2388  & 0.0233 \\
ResNet34       & 0.1462  & 0.1149  & 0.0226 \\
ResNet50       & 0.1445  & 0.1145  & 0.0221 \\
\bottomrule
\end{tabular}
\caption*{(a) Solar Wind Forecasting}
\end{minipage}%
\hfill
\begin{minipage}{0.48\textwidth}
\centering
\begin{tabular}{lcccc}
\toprule
\textbf{Model} & \textbf{TSS} & \textbf{HSS} & \textbf{CSS} & \textbf{F1} \\
\midrule
AlexNet    & 0.359 & 0.354 & 0.356 & 0.679 \\
MobileNet  & 0.326 & 0.312 & 0.319 & 0.662 \\
ResNet18   & 0.320 & 0.317 & 0.318 & 0.660 \\
ResNet34   & 0.290 & 0.289 & 0.289 & 0.645 \\
ResNet50   & 0.261 & 0.281 & 0.271 & 0.627 \\
\bottomrule
\end{tabular}
\caption*{(b) Solar Flare Prediction}
\end{minipage}

\label{tab:baseline-performance}
\end{table}

\begin{table}[H]
\caption{{Baseline performance for (a) EVE Prediction and (b) AR Emergence Forecasting on test data}}

\begin{minipage}{0.48\textwidth}
\centering
\begin{tabular}{lccc}
\toprule
\textbf{Model} & \textbf{RMSE} & \textbf{MAE} & \textbf{Val Loss} \\
\midrule
UNet           & 0.0754  & 0.0558  & 0.00569 \\
AttentionUNet  & 0.0754  & 0.0558  & 0.00569 \\
ResNet18       & 0.2108  & 0.2388  & 0.02330 \\
ResNet34       & 0.1462  & 0.1149  & 0.02255 \\
ResNet50       & 0.1445  & 0.1145  & 0.02208 \\
\bottomrule
\end{tabular}
\caption*{(a) EVE Prediction}
\end{minipage}
\hfill
\begin{minipage}{0.48\textwidth}
\centering
\begin{tabular}{lcc}
\toprule
\textbf{Model} & \textbf{MSE} & \textbf{RMSE} \\
\midrule
ST Attention & 0.1538 & 0.3921 \\
ST ResNet    & 0.1527 & 0.3908 \\
(LSTM)    & 0.0140 & 0.1180 \\
\bottomrule
\end{tabular}
\caption*{(b) AR Emergence Forecasting (ST: Spatiotemporal)}
\end{minipage}
\end{table}


\section{Limitations}
\label{sec:limitations}
Solar data have a few important features that must be taken into consideration:
\begin{itemize}[leftmargin=*]
    \item \textbf{\textit{Solar rotation:}} The Sun takes $\approx27$ days to complete rotation, also referred to as a Carrington rotation. This results in a repeat of magnetic structures every $\approx27$ days~\cite{owens201327}. It is preferable to separate the training and testing sets by at least 1/2 Carrington rotation to avoid observing the repeating spatial patterns.
    The standard practice in heliophysics is to use temporally non-overlapping training-validation-testing splits \cite{Pandey2021} (e.g., first 5 years for training, next 3 years for validation, and remaining years for testing, or the first 8 months of each year for training (Jan-Aug), the next two months (Sep-Oct) for validation, and the last two (Nov-Dec) for testing.)
    \item \textbf{\textit{Solar Cycle:}} Solar activity undergoes an $\approx11$ year maximum and minimum. This results in a subtle, 11-year variation in the solar activity~\cite{schwenn2007solar}. Hence, it is ideal to perform data splitting by sampling activity across the whole solar cycle.  The standard practice in heliophysics is to maximize dataset coverage to contain at least a full solar cycle (e.g., 2010-2022).  This in combination with the split mentioned above, ensures the creation of representative training-validation-test splits.
    \item \textbf{\textit{Ecliptic angle:}}  The plane of the Earth's orbit is inclined with respect to the equator.  Because of this, as the year progresses, the Earth goes slightly above (below) the north (south) pole.  While our data is fixed so that the solar north is always pointing upwards, the solar disk center is almost never on the equator.  Any application aiming to use heliographic coordiantes (i.e. latitude and longitude on the solar surface), must account for this perspective effect of the Earth's orbit.
    \item \textbf{\textit{Lack of farside observations:}}  While SDO provides one of the most comprehensive datasets, the observations are limited to the visible (Earth) side of the Sun. Many of the applications using this dataset can be directly impacted by events occurring on the farside of the Sun.
\end{itemize}

\section{Code Availability} 

The SuryaBench datasets are publicly available on Huggingface: \url{https://huggingface.co/collections/nasa-impact/suryabench-68265ce306fc2470c121af7b}. Our code for dataset preparation and creation, and baseline model training is publicly available at  \url{https://github.com/NASA-IMPACT/SuryaBench}.

\bibliography{references}

\section{Author Contributions}

Sujit Roy: Conceptualization, Methodology, Data production, Visualization, Writing–original draft, Writing–review \& editing, Project administration.\\  
Johannes Schmude: Conceptualization, Methodology, Baseline design, Data Validation.  \\ 
Vishal Gaur: Methodology, Data production, Visualization, Writing–original draft, Baseline design.  
Rohit Lal: Methodology, Data production, Visualization, Writing–original draft, Baseline design.  \\ 
Dinesha V. Hegde: Data preprocessing and production (SDO ML Ready), Visualization, Writing–original draft, Writing–review \& editing.  \\ 
Amy Lin: Data production, Visualization.  \\ 
Talwinder Singh: Data production (Active region segmentation and flare forecasting), Visualization.  \\ 
Berkay Aydin: Data production (Active region segmentation and flare forecasting), Writing–editing original draft.  \\
Andrés Muñoz-Jaramillo: Data production coordination for application benchmark datasets, Writing–original draft.  \\ 
Vishal Upendran: Data production (solar wind), editing - original draft  \\ 
Daniel da Silva: Data production (Coronal field extrapolation), editing-original draft.  \\ 
Shah Bahaudding: Data production (Solar EUV spectra).  \\ 
Spiridon Kasapis: Data production (Active Region Emergence forecast)
Kang Yang: Baseline design for AR segmentation.  \\ 
Chetraj Pandey: Baseline design for flare forecasting.  \\ 
Iksha Gurung: Data Hosting \\
Jinsu Hong: Baseline design and experiments for flare forecasting and AR segmentation.\\ 
Nikolai Pogorelov: Validation \& Review \\
Manil Maskey: Validation \& Review \\
Rahul Ramachandran: Validation, Review \& Project administration.\\
\section{Competing Interests}
The authors declare no competing interests.

\section{Acknowledgements}
This work is supported by NASA Grant 80MSFC22M004. The Authors acknowledge the National Artificial Intelligence Research Resource (NAIRR) Pilot and NVIDIA for providing support under grant no. NAIRR240178.

\clearpage
\section*{Supplementary Information}

\subsection*{Low Quality AIA Data and QUALITY Keyword}

\begin{figure}[!h]
\centering
\includegraphics[width=0.9\textwidth]{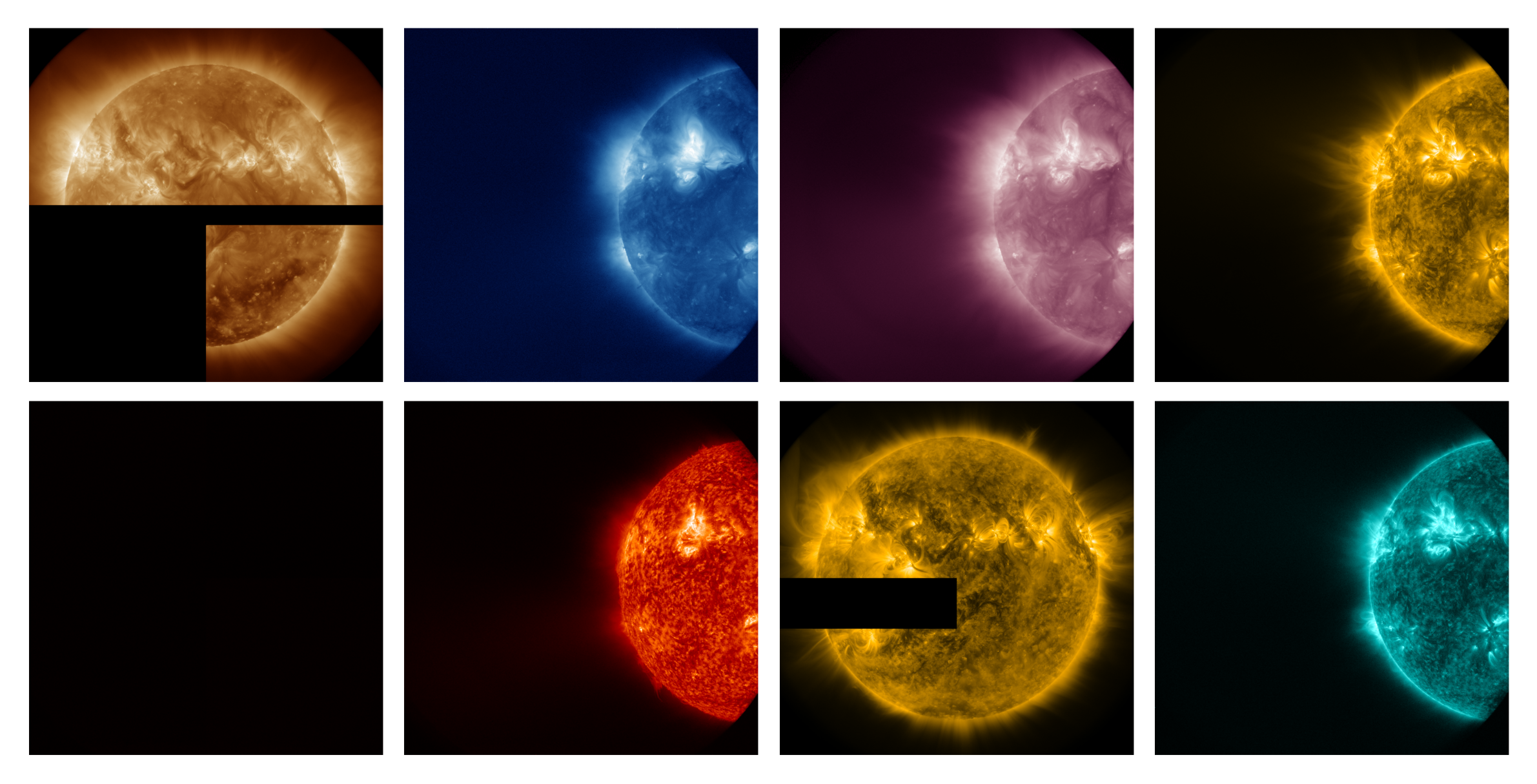} 
\caption{Bad AIA measurements due to a variety of reasons.}
\label{fig:baddata}
\end{figure}

AIA image headers and data may be affected by operational events such as off-pointing, defocusing, or missing data during eclipse seasons. These issues are flagged by a non-zero QUALITY keyword in the image header. It is important to check the QUALITY before detailed analysis of AIA data. Note that the QUALITY keyword is a 32-bit integer with bitwise flags. We present a set of examples in Figure~\ref{fig:baddata}.

\clearpage

\subsection*{Overview of Baseline Learning Models for Application Benchmark Datasets}

This supplementary material provides a detailed examination of the  tasks from a machine learning perspective, including task-specific objectives, model architectures, and relevant implementation details.

\subsubsection*{Active region emergence forecast}

We address the task of \textbf{predicting continuum intensity} over a spatially-distributed grid of solar active regions, using historical measurements of solar magnetic flux and acoustic power. This task encapsulates a complex spatiotemporal forecasting problem grounded in heliophysics, where both \textbf{temporal dependencies} and \textbf{spatial interactions} are crucial for accurate modeling. Understanding and forecasting continuum intensity has strong implications for solar physics and space weather prediction. High-intensity regions on the solar surface are often precursors to flare activity and magnetic storms.

\begin{figure}[!b]
\centering
\includegraphics[width=\textwidth]{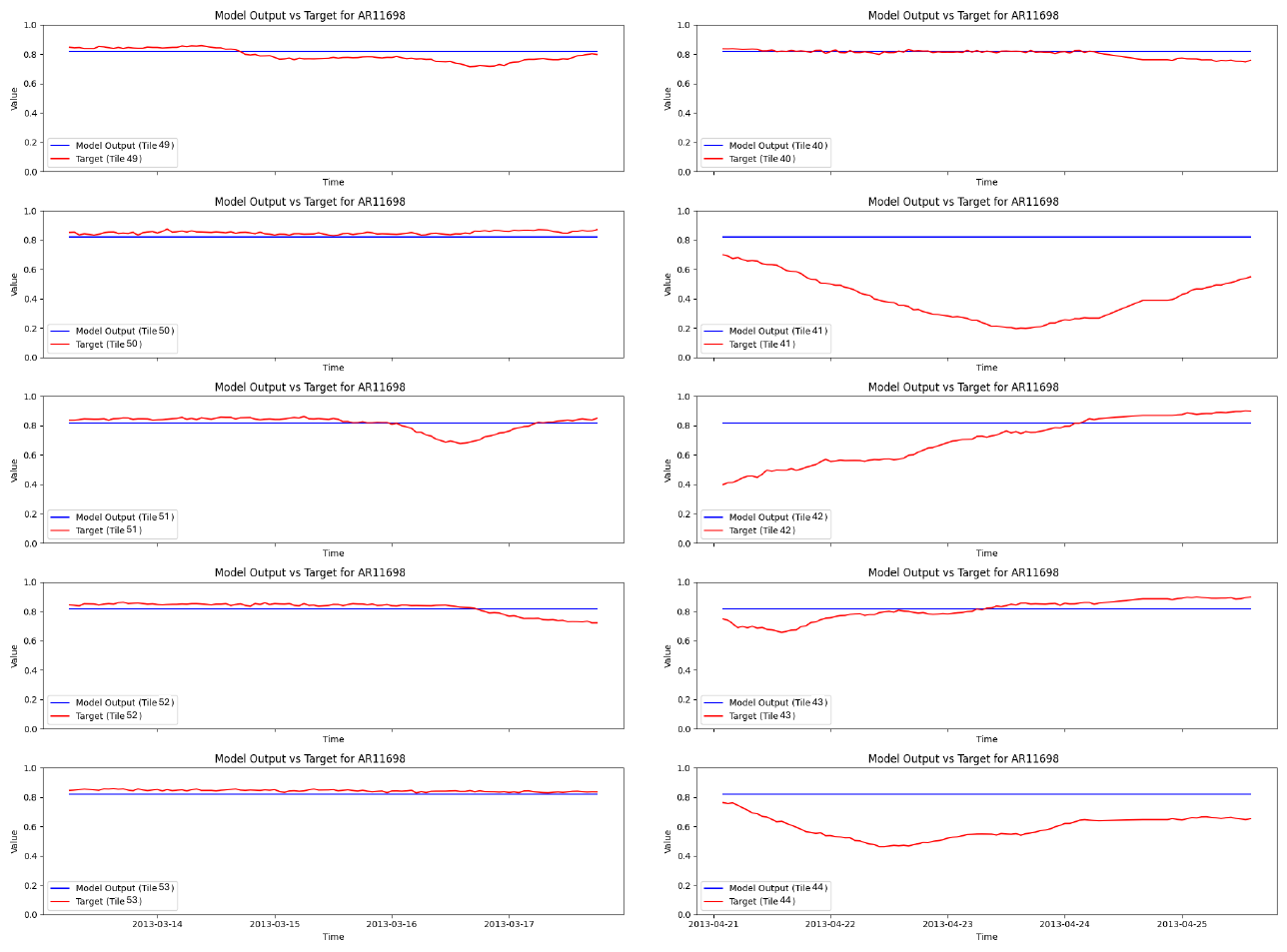}
\caption{Results for AR emergence forecasting using the HelioFM when validating on AR11698 (left) and AR11726 (right).}
\label{fig:results11698}
\end{figure}

\noindent\textbf{Problem Formulation: }Given a tracked solar active region observed over a sequence of $T = 120$ time steps (i.e., 1 day in our dataset), our objective is to predict the \textit{continuum intensity} at a subsequent time point for each spatial grid cell. Formally, let:

\begin{itemize}
    \item $\mathbf{X} \in \mathbb{R}^{T \times C \times S}$ denote the input tensor for a region, where:
    \begin{itemize}
        \item $T = 120$ is the number of time steps (\textasciitilde24 hours, sampled at 12-minute cadence),
        \item $C = 5$ is the number of physical quantities per cell,
        \item $S = 63$ is the number of spatial cells in the tracked region.
    \end{itemize}
    \item $\mathbf{y} \in \mathbb{R}^{S}$ denote the target continuum intensity per cell.
\end{itemize}

The model is trained to approximate a function $f_\theta: \mathbb{R}^{T \times C \times S} \rightarrow \mathbb{R}^{S}$ that maps the temporal and spatial patterns of input features to the scalar output intensities.

The five input channels correspond to two types of physical measurements:

\begin{itemize}
    \item \textbf{Mean Unsigned Magnetic Flux (1 channel)}: captures the net strength of local magnetic fields in the region.
    \item \textbf{Doppler Velocity Acoustic Power (4 channels)}: measured across four distinct frequency bands: 2--3, 3--4, 4--5, and 5--6 mHz, these capture multi-scale oscillatory dynamics linked to wave propagation and subsurface flows.
\end{itemize}

Spatially, each active region is tracked and cropped into a 9\texttimes9 grid of tiles. However, we discard the top and bottom rows for normalization reasons (e.g., to suppress edge artifacts), resulting in a 7\texttimes9 = 63 spatial cells. For each time step, the full tensor $\mathbf{X}_t \in \mathbb{R}^{C \times S}$ captures these 5-channel inputs over the grid.

\noindent\textbf{Dataset and Temporal Context: }
The dataset comprises 3,479 unique regions, indexed and temporally aligned. For each region:

\begin{itemize}
    \item \textbf{Input timestamps} span a window of 120 steps (\textasciitilde24 hours),
    \item \textbf{Output timestamp} is a single future point for which continuum intensity is predicted.
\end{itemize}

\noindent\textbf{Evaluated Model Architectures}

\begin{figure*}[!h]
    \centering
    \begin{subfigure}[b]{0.45\linewidth}
        \centering
        \includegraphics[width=0.7\linewidth]{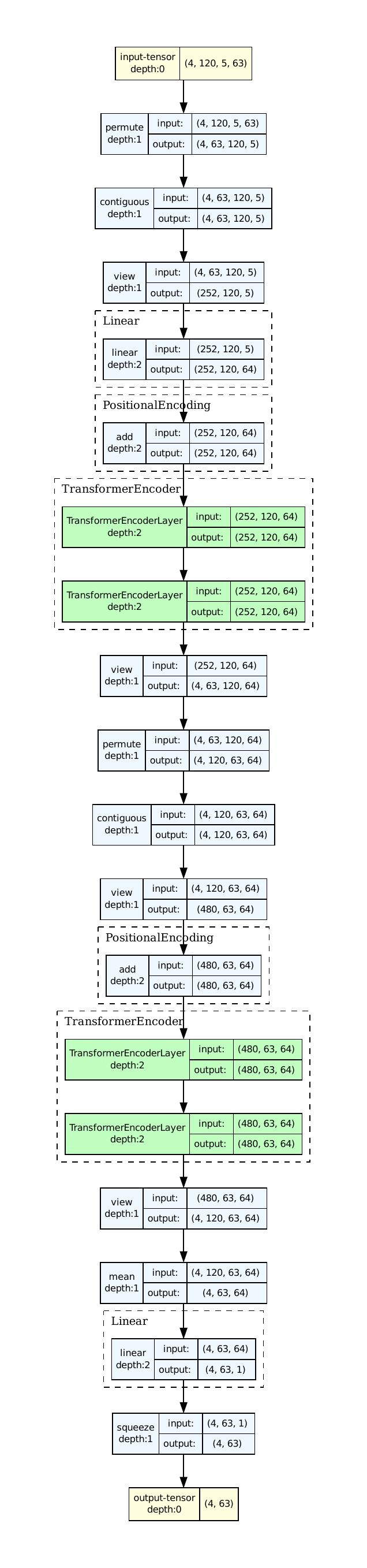}
        \caption{SpatioTemporal Attention}
        \label{fig:spatiotemporal-attention}
    \end{subfigure}
    \hfill
    \begin{subfigure}[b]{0.45\linewidth}
        \centering
        \includegraphics[width=0.7\linewidth]{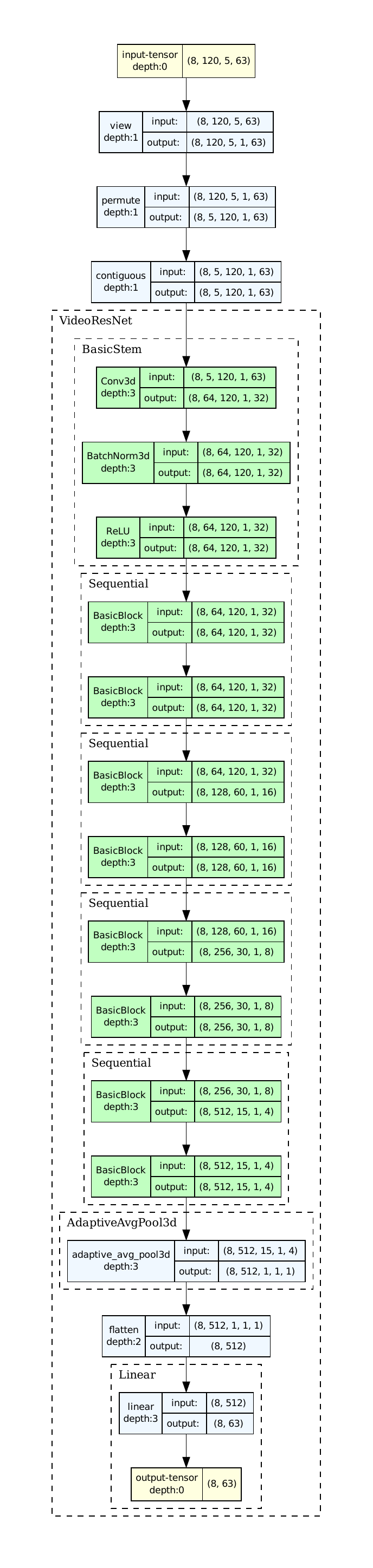}
        \caption{SpatioTemporal ResNet}
        \label{fig:spatiotemporal-resnet}
    \end{subfigure}
    \caption{Comparison between the proposed \textbf{SpatioTemporalAttention} model and a baseline \textbf{SpatioTemporal ResNet}. The SpatioTemporalAttention model explicitly decomposes the modeling task into two sequential Transformer stages: temporal attention applied independently to each spatial grid cell, followed by spatial attention across all cells at each timestep. In contrast, the SpatioTemporal ResNet baseline uses a 3D convolutional backbone adapted from \texttt{r3d\_18} to learn spatiotemporal features jointly through hierarchical convolutional filters.}    \label{fig:side-by-side-models}
\end{figure*}

We explore two complementary spatiotemporal modeling paradigms: SpatioTemporal Transformer and SpatioTemporal ResNet. The SpatioTemporal Transformer model is a two-stage Transformer that sequentially models temporal and spatial dynamics. Its structure reflects a deliberate architectural bias aligned with the problem's domain priors:

\begin{enumerate}
    \item \textbf{Temporal Attention (per cell)}: For each of the 63 spatial grid cells, we treat the 120-step sequence of 5-channel inputs as a time series. These are projected to a $d = 64$-dimensional embedding and passed through a Transformer encoder with positional encodings, allowing the model to learn temporal dependencies in each spatial location independently.
    \item \textbf{Spatial Attention (per timestep)}: At each of the 120 timesteps, the spatial pattern of cell embeddings is modeled as a sequence of 63 tokens. A second Transformer block captures interactions and correlations between different locations on the solar disk.
    \item \textbf{Output Head}: After pooling over time, the spatial embeddings are passed through a linear regressor to predict a scalar continuum intensity for each cell.
\end{enumerate}

This modular design allows the model to explicitly factorize temporal and spatial reasoning, which is beneficial for interpretability and transfer across regions with similar temporal but different spatial dynamics.

As a second baseline, we also implement a 3D convolutional model based on the ResNet-18 video backbone (\texttt{r3d\_18}), i.e., SpatioTemporal ResNet. Here, the input tensor is reshaped to match the expected input for Conv3D networks:

\begin{itemize}
    \item $\mathbf{X} \rightarrow \mathbb{R}^{B \times C \times T \times H \times W}$  
    with $H = 1$, $W = 63$, and $C = 5$.
\end{itemize}

This network is initialized with pretrained weights (optional), and the first convolutional layer is modified to accept 5 input channels instead of 3 (RGB). The final fully connected layer is replaced to regress 63 outputs. This model captures hierarchical spatiotemporal correlations through convolutional filters, offering a computationally efficient and generalizable baseline.


\noindent\textbf{Results and Observations:} Models are evaluated using mean squared error (MSE) and mean absolute error (MAE) over all 63 spatial grid cells, averaged across held-out validation regions. Since outputs are per-cell continuous intensities, these metrics offer a direct measure of spatial forecasting accuracy.


\subsubsection*{Solar Flare Forecasting}

Solar flare forecasting is framed as a binary classification task where the goal is to predict whether a strong solar flare (i.e., M- or X-class) will occur within a 24-hour window following a given observation time $t$. The prediction window spans the interval $[t, t + 24)$ hours. Within this window, multiple solar flare events may occur. To assign a label to the input at time $t$, two labeling strategies are used:

\begin{itemize}
    \item \textbf{Maximum Flare Intensity}: We select the flare with the highest intensity in the 24-hour prediction window. If this maximum intensity exceeds a threshold of $10^{-4}$ W/m\textsuperscript{2}, the input is labeled as positive (flare will occur); otherwise, it is labeled negative.
    \item \textbf{Cumulative Flare Intensity}: We sum the intensities of all flares that occur in the prediction window. If the cumulative intensity exceeds a threshold of $10$, the input is labeled positive; otherwise, negative.
\end{itemize}

\noindent\textbf{Problem Formulation: } Given solar observation data at time \( t \), the goal is to predict whether a significant solar flare will occur within the subsequent 24-hour window, i.e., in the interval \([t, t + 24)\). This is framed as a binary classification task:

\[
f(\mathbf{x}_t) \rightarrow \{0, 1\}
\]

where \( \mathbf{x}_t \in \mathbb{R}^{C \times H \times W} \) (or potentially \( \mathbb{R}^{T \times C \times H \times W} \) for temporal stacks) represents the multi-channel input image (or sequence) at time \( t \), and the output is a binary label:
\[
y_t =
\begin{cases}
1, & \text{if a flare is expected in } [t, t+24) \\
0, & \text{otherwise}
\end{cases}
\]

To determine \( y_t \), two flare labeling strategies are considered:

\begin{itemize}
    \item \textbf{Maximum Flare Intensity:} The label is set to 1 if the maximum flare intensity in \([t, t+24)\) exceeds a fixed threshold \( \theta_{\text{max}} = 10^{-4} \) W/m\textsuperscript{2}.
    \item \textbf{Cumulative Flare Intensity:} The label is set to 1 if the sum of all flare intensities in \([t, t+24)\) exceeds a threshold \( \theta_{\text{sum}} = 10 \).
\end{itemize}

\noindent\textbf{Evaluated Model Architectures:} We evaluate the performance of several standard convolutional neural network (CNN) architectures adapted for binary classification. Each model takes in spatial or spatiotemporal representations of solar magnetic field or other physical parameters (details omitted here) and outputs a binary prediction. Evaluated architectures include the following: 
\begin{itemize}
    \item \textbf{AlexNet}: A lightweight CNN with five convolutional layers followed by three fully connected layers. Its shallow depth makes it faster to train and less prone to overfitting in small datasets.
    \item \textbf{MobileNet}: A mobile-optimized architecture using depthwise separable convolutions to reduce computational cost. Useful for efficient forecasting on edge devices.
    \item \textbf{ResNet18 / ResNet34 / ResNet50}: Residual Networks with varying depths (18, 34, and 50 layers respectively), incorporating skip connections to enable better gradient flow and deeper representations.
\end{itemize}

All models are modified with a final fully connected layer followed by a sigmoid activation to output a probability score for binary classification.

\noindent\textbf{Results and Obsersations:} Table~\ref{tab:flare_results} shows the evaluation metrics we used for solar flare forecasting:
\begin{itemize}
    \item \textbf{TSS (True Skill Statistic)}: Measures the model's ability to distinguish between flare and non-flare events.
    \item \textbf{HSS (Heidke Skill Score)}: Accounts for both hits and false alarms.
    \item \textbf{CSS (Composite Skill Score)}: Provides a balanced measure as the geometric mean of TSS and HSS.
    \item \textbf{F1 Score}: Harmonic mean of precision and recall.
\end{itemize}

\begin{table}[h]
\centering
\begin{tabular}{lcccc}
\toprule
\textbf{Model} & \textbf{TSS} & \textbf{HSS} & \textbf{CSS} & \textbf{F1} \\
\midrule
AlexNet   & 0.359 & 0.354 & 0.356 & 0.679 \\
MobileNet & 0.326 & 0.312 & 0.319 & 0.662 \\
ResNet18  & 0.320 & 0.317 & 0.318 & 0.660 \\
ResNet34  & 0.290 & 0.289 & 0.289 & 0.645 \\
ResNet50  & 0.261 & 0.281 & 0.271 & 0.627 \\
\bottomrule
\end{tabular}
\caption{Solar Flare Forecasting Performance}
\label{tab:flare_results}
\end{table}

\begin{itemize}
    \item \textbf{TSS (True Skill Statistic)}:
    \[
    \mathrm{TSS} = \frac{TP}{TP + FN} - \frac{FP}{FP + TN}
    \]
    Measures the ability to distinguish between flare and non-flare events. Ranges from -1 (inverse prediction) to +1 (perfect prediction), with 0 indicating no skill.

    \item \textbf{HSS (Heidke Skill Score)}:
    \[
    \mathrm{HSS} = \frac{2(TP \cdot TN - FP \cdot FN)}{(TP + FN)(FN + TN) + (TP + FP)(FP + TN)}
    \]
    Evaluates performance relative to random chance, considering both hits and false alarms. Ranges from \(-\infty\) to 1.

    \item \textbf{CSS (Composite Skill Score)}:
    \[
CSS = 
\begin{cases}
0, & \text{if } TSS < 0  \text{ OR } HSS < 0 \\
\sqrt{TSS \times HSS}, & \text{otherwise}
\end{cases}    \]
    It measures the geometric mean of TSS and HSS when they are positive.

    \item \textbf{F1 Score}:
    \[
    \mathrm{F1} = \frac{2 \cdot \text{Precision} \cdot \text{Recall}}{\text{Precision} + \text{Recall}} = \frac{2TP}{2TP + FP + FN}
    \]
    Harmonic mean of precision and recall, balancing both false positives and false negatives.
\end{itemize}

The results indicate that AlexNet performs best among the evaluated architectures across all metrics, potentially due to its shallower structure and better generalization on limited data. Deeper architectures such as ResNet50 may suffer from overfitting or excessive capacity relative to the dataset size.

\subsubsection*{Solar Wind Forecasting}

Solar wind forecasting is a critical regression task aimed at predicting the solar wind speed at a given spatial point, specifically within a prediction window of 4 days following an observation time $t$. Precise forecasting of solar wind speeds is fundamental for mitigating the adverse effects of space weather on satellite communication systems, navigation systems, and electrical grids on Earth.

This dataset comprises scalar measurements of solar wind speeds, recorded hourly from 2010-01-01 through 2023-12-31, resulting in a temporally rich dataset with substantial coverage of solar cycles. The solar wind speed values exhibit significant variability, ranging from $2.4 \times 10^{2}$ km/s to $8.8 \times 10^{2}$ km/s.

\noindent\textbf{Problem Formulation}

Formally, given solar observation data (such as AIA and HMI multi-channel solar imaging data) represented by $\mathbf{x}_t$ at observation time $t$, the task is to predict the scalar solar wind speed at time $t + \Delta t$, where $\Delta t = 4$ days:

$$
y_{t+\Delta t} = f(\mathbf{x}_t),
$$

where $\mathbf{x}_t \in \mathbb{R}^{C \times H \times W}$ represents the multi-channel, high-resolution input imagery data at time $t$, and $y_{t+\Delta t} \in \mathbb{R}$ represents the predicted scalar solar wind speed.

\begin{figure*}[!htpb]
    \centering
    \includegraphics[width=0.7\linewidth]{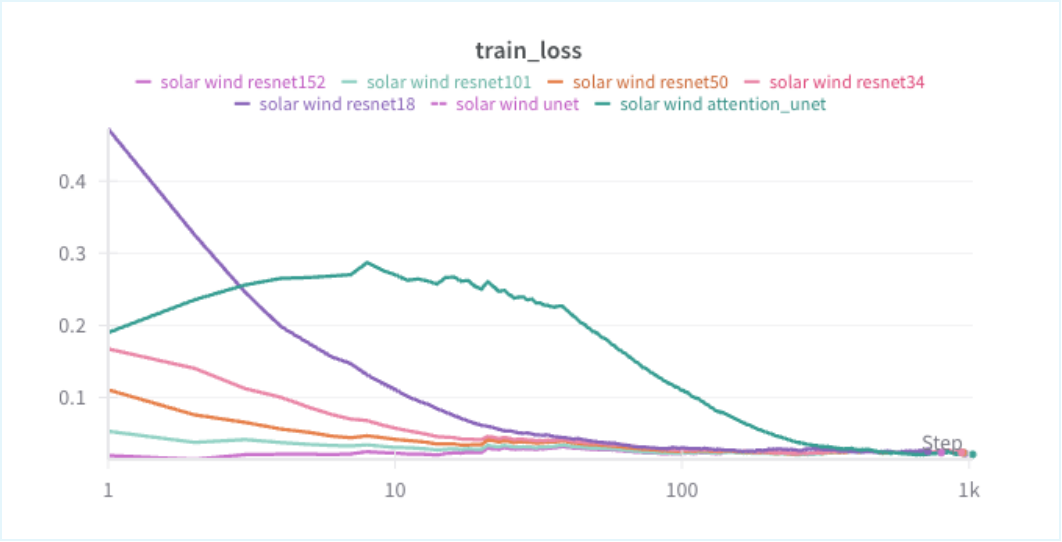}
    \caption{Training loss curves for various deep learning architectures on the solar wind forecasting task. The comparison includes ResNet variants (18, 34, 50, 101, 152), U-Net, and Attention U-Net. Models with deeper architectures (e.g., ResNet152) and attention mechanisms (e.g., Attention U-Net) tend to converge faster and reach lower final training losses, indicating their superior capacity to model the input-output mapping for this regression task. Log-scale on the x-axis highlights early training dynamics.}
    \label{fig:loss_swc}
\end{figure*}

\begin{figure*}
    \centering
    \begin{subfigure}[b]{0.45\linewidth}
        \centering
        \includegraphics[width=0.5\linewidth]{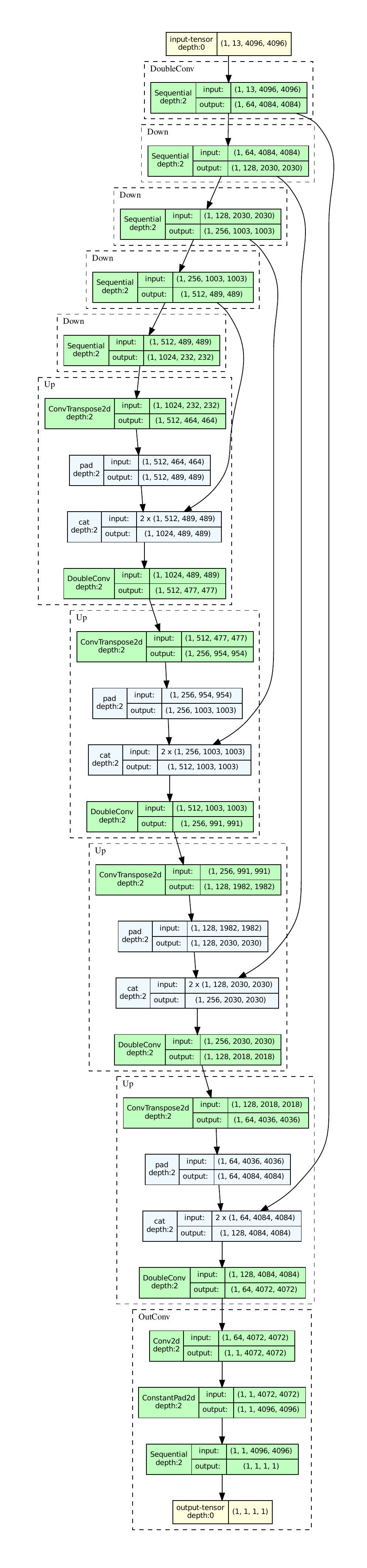}
        \caption{UNet Model}
        \label{fig:unet}
    \end{subfigure}
    \hfill
    \begin{subfigure}[b]{0.45\linewidth}
        \centering
        \includegraphics[width=0.5\linewidth]{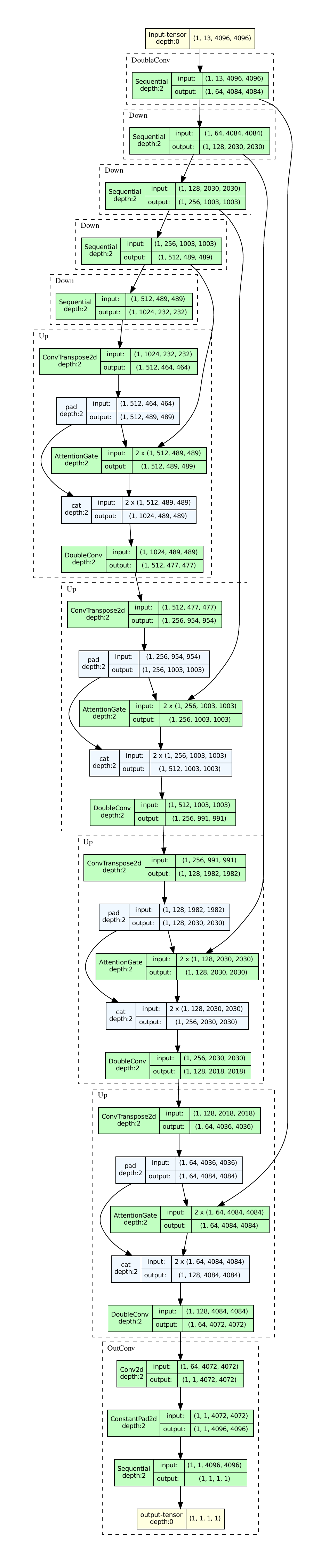}
        \caption{Attention Unet Model}
        \label{fig:attnUnet}
    \end{subfigure}
    \caption{Detailed architectural diagrams of (a) the UNet model and (b) the Attention UNet model used for predicting solar EUV irradiance and solar wind. The Attention UNet enhances the standard UNet architecture by incorporating attention gates, allowing the network to selectively emphasize relevant spatial features and improve predictive performance in complex regression tasks such as high-dimensional spectral prediction.}    \label{fig:unet-side-by-side-models_sw_shah}
\end{figure*}

\noindent\textbf{Evaluated Model Architectures:}Figure \ref{fig:unet-side-by-side-models_sw_shah} illustrates the architectures utilized for the solar wind forecasting task, emphasizing the distinction between attention-driven and baseline UNet approaches.

We explored and benchmarked several state-of-the-art deep learning architectures detailed below:
\begin{itemize}
    \item \textbf{Attention UNet:} The Attention UNet architecture enhances the traditional UNet through the integration of attention gates, facilitating dynamic suppression of irrelevant spatial features and emphasizing salient regions pertinent to predicting solar wind characteristics. Attention gates, introduced within skip connections between encoder and decoder paths, adaptively weigh encoder outputs based on decoder contexts, significantly improving the discriminative capability of the model. Given the large resolution (4096 $\times$ 4096 pixels) of solar imagery data, adaptive average pooling followed by convolutional layers was strategically employed to condense feature representations, subsequently enabling precise regression to the scalar solar wind speed.

\item \textbf{Standard UNet:} A standard UNet architecture served as a robust baseline, providing a fundamental encoder-decoder structure with straightforward skip connections. Its primary role was to gauge the incremental benefit derived from attention mechanisms explicitly integrated into the Attention UNet model.

\item \textbf{ResNet-based Convolutional Models:} As an additional comparative baseline, we employed ResNet architectures (ResNet-18, ResNet-34, and ResNet-50) to leverage deep residual learning's inherent capabilities in capturing complex hierarchical features. These models were initially pre-trained (optional) and specifically adapted for solar data by modifying the first convolutional layer to accept 13 input data representative of solar observational channels (instead of the standard RGB inputs). The final layer of the network was adapted to produce a single scalar output directly.

\end{itemize}

Notably, we achieved optimal performance metrics and lowest validation loss with the ResNet-50 architecture, likely attributed to its deeper structure and larger parameter count (33 million parameters), facilitating richer representation of complex spatiotemporal solar dynamics.

\noindent\textbf{Results and Observations:} Our experiments indicate that incorporating attention mechanisms (Attention UNet) improves the predictive performance over standard UNet, highlighting the importance of adaptive feature weighting in solar wind forecasting. Additionally, deeper architectures such as ResNet-50 outperform shallower networks, emphasizing the complexity and depth required to model solar physics phenomena effectively. These observations underscore a fundamental insight: architectural complexity and context-aware feature selection are critical components in accurately predicting space weather events. Training losses for this  task can be found in Figure \ref{fig:loss_swc}.

\subsubsection*{ Solar EUV spectra prediction}

Predicting solar Extreme Ultraviolet (EUV) irradiance accurately is crucial for understanding and forecasting space weather, which directly impacts satellite operations, communication systems, and navigation. This task involves forecasting irradiance across a spectrum of 1343 spectral channels, reflecting complex spatial and temporal patterns captured by solar imagery.

Formally, we frame this as a regression problem: Given multi-channel solar imaging data $\mathbf{x}_t \in \mathbb{R}^{C \times H \times W}$ at time $t$, the goal is to predict a continuous vector of EUV irradiance values $y_t \in \mathbb{R}^{1343}$:

$$y_t = f(\mathbf{x}_t)$$

\noindent\textbf{Evaluated Model Architectures:} To establish baselines for this task, we evaluated several deep learning architectures commonly used in computer vision and scientific regression tasks. Table 4(a) for main paper presents the performance of these baseline models, comparing both convolutional networks (ResNet variants) and segmentation-inspired architectures (U-Net and Attention U-Net).

\noindent\textbf{Results and Observations:}
The performance metrics employed include Root Mean Square Error (RMSE), Mean Absolute Error (MAE), and validation loss. These metrics provide complementary views of predictive accuracy and model robustness. U-Net and Attention U-Net significantly outperform traditional convolutional networks, underscoring the efficacy of architectures that inherently model spatial correlations and multi-scale features in predicting complex, high-dimensional irradiance spectra.

This predictive modeling task not only benchmarks model capabilities in handling high-dimensional regression but also advances the applicability of deep learning methods to critical solar physics-driven applications.

\begin{figure*}[!h]
    \centering
    \includegraphics[width=0.7\linewidth]{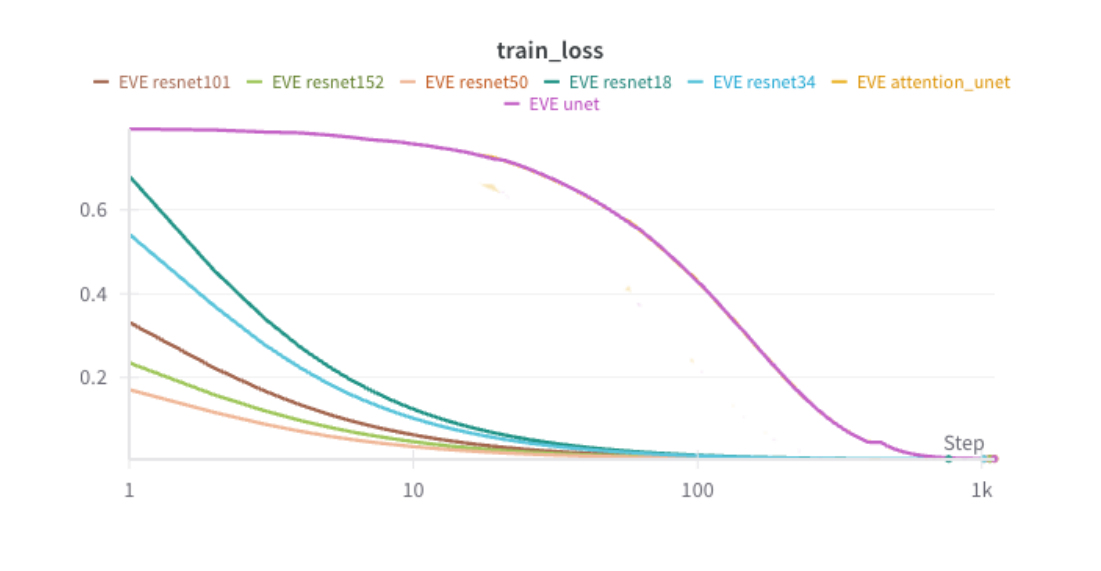}
    \caption{Training loss curves comparing various baseline models (ResNet18, ResNet34, ResNet50, ResNet101, ResNet152, UNet, and Attention UNet) for the Solar EUV irradiance prediction task. UNet and Attention UNet show slower initial convergence compared to ResNet variants but achieve significantly lower final loss values, highlighting their effectiveness in modeling complex spatial patterns inherent to solar EUV spectra.}
    \label{fig:loss_sw}
\end{figure*}

\end{document}